\documentclass[10pt]{article}
\usepackage[margin=1in]{geometry}

\usepackage[english]{babel}

\usepackage[margin=1in]{geometry}

\usepackage{amsthm}

\usepackage{amsmath} 
\usepackage{float}
\usepackage{graphicx}
\usepackage{pdflscape}
\usepackage{amsfonts}
\usepackage{subcaption}
\usepackage{xcolor,colortbl}
\usepackage[hidelinks]{hyperref}
\hypersetup{
	colorlinks  = true,
	linkcolor  = {red!100!black},
	citecolor  = {blue!100!black},
	urlcolor     = {blue!100!black}
}
\usepackage{tabularx}
\usepackage{multirow}
\usepackage{algorithm}
\usepackage{algpseudocode}
\usepackage{booktabs}
\usepackage{nameref}
\usepackage{comment}
\usepackage{caption}
\usepackage[utf8]{inputenc}
\usepackage[english]{babel}
\newtheorem{Theorem}{Theorem}
\newtheorem{Proposition}{Proposition}
\newtheorem{Corollary}{Corollary}

\newtheorem{Remark}{Remark}
\newtheorem{Assumption}{Assumption}

\title{A Discrete-time Dynamical Model for Optimal Dispatching and Rebalancing of Autonomous Mobility-on-Demand Systems}
\author{Ali~Aalipour
        and~Alireza Khani
\thanks{
Ali Aalipour is with the Electrical and Computer Engineering department, University of Minnesota, MN, USA {\tt\small aalip002@umn.edu}.
\\
Alireza Khani is with the Civil, Environmental, and Geo-Engineering department, University of Minnesota, MN, USA {\tt\small akhani@umn.edu}.
}
}
\begin{document}
\maketitle

\begin{abstract}
Autonomous vehicles are rapidly evolving and will soon enable the application of large-scale mobility-on-demand (MoD) systems. Managing the fleets of available vehicles, commonly known as "rebalancing," is crucial to ensure that vehicles are distributed properly to meet customer demands. This paper presents an optimal control approach to optimize vehicle scheduling and rebalancing in an autonomous mobility-on-demand (AMoD) system. We use graph theory to model a city partitioned into virtual zones. Zones represent small areas of the city where vehicles can stop and pick up/drop off customers, whereas links denote corridors of the city along which autonomous vehicles can move. They are considered vertices and edges in the graph. Vehicles employed in the AMoD scheme are autonomous, and rebalancing can be executed by dispatching available empty vehicles to areas undersupplied.
Rebalancing is performed on the graph's vertices, i.e., between city areas. We propose a linear, discrete-time model of an AMoD system using a transformed network. After acquiring the model, the desired number of rebalancing vehicles for the AMoD model is derived through an optimization problem. Moreover, the well-posedness of the model is illustrated. To leverage the proposed model, we implemented the model predictive control (MPC) framework to find the optimal rebalancing and scheduling policy. We show the MPC's effectiveness and how the MPC framework can be implemented in real-time for a real-world case study.
The numerical results show that the MPC with a linear cost function and linear reference, which it tracks, is effective, outperforming other MPC-based and state-of-the-art algorithms across all evaluation criteria.

\end{abstract}

\section{Introduction}
Nowadays, people are able to use MoD services to travel and share vehicles with other people by sending requests through mobile devices. A MoD can be replaced by AMoD due to the lower labor costs of autonomous vehicle (AV) operations. The cooperative nature of AVs is in contrast with selfish taxi drivers seeking to maximize their profits. By optimizing routing, rebalancing, charging schedules, etc., central coordination can minimize externalities in AMoD systems. Further, customers do not have to drive, which allows them to save time commuting. Various companies started to develop this technology in response to such promising benefits.

Fleet management studies focus on optimizing vehicle routing to rebalance empty vehicles and serve the customers in the network. They aim to reduce operational costs and waiting times for customers. The AMoD system avoids the costs of rebalancing drivers to drive vehicles from oversupplied origins to undersupplied origins. Moreover, similar to current car-sharing companies such as Car2go, AMoD provides excellent convenience for single-way trips since the users do not have to return vehicles to the origin of the trip. AMoD services potentially offer a good opportunity for efficient fleet management. 

Due to recent developments in technology and customer preferences, AMoD systems have drawn a significant amount of attention, and an in-depth study on many various aspects of the AMoD system is currently underway.
As some origins and destinations are more popular than others, AMoD systems frequently become unbalanced; AVs accumulate at some stations. So, available empty AVs are rebalancing the network by serving the undersupplied regions. Designing the rebalancing policy has been studied in 
\cite{pavone2012robotic}, \cite{volkov2012markov}, and \cite{acquaviva2014novel}. 

\subsection{Literature Review}
At a high level, approaches to tackling operational AMoD issues may be essentially divided into two categories: model-free and model-based.
\subsubsection{Model-free}
Model-free techniques for AMoD fleet rebalancing can be characterized as centralized or decentralized. A centralized agent rebalances cars in order to optimize specific objectives, such as travel time. 
The authors in \cite{turan2020dynamic} present a RL approach that uses a dynamic pricing autonomous MoD framework aiming to maximize the profit and rebalance the fleet. 
They focus on vehicle charging and discharging policies, whereas \cite{skor2021modular} emphasizes customer waiting time minimization.
A RL approach for the taxi dispatching and rebalancing problem is introduced in \cite{gao2018optimize} to maximize taxi driver long-term revenues utilizing Q-learning and discrete states and actions on a grid-shaped map. A double deep Q-learning architecture is proposed in \cite{guo2022deep} for vehicle routing in a ride-sharing AMoD system, where idle cars are rebalanced to fulfill future demands. \cite{fluri2019learning} employs RL and mixed integer linear programming (MILP) for fleet rebalancing and uses hierarchical binary partitioning and tabular Q-learning for RL.

Decentralized approaches, as contrasted with centralized methods, enable each vehicle to act as its own agent and be trained in either a cooperative or competitive fashion. In \cite{al2019deeppool}, proposes a ride-sharing architecture for vehicles utilizing a deep Q-network to learn the optimal policies for individual vehicles in a distributed and uncoordinated fashion. Passenger satisfaction and vehicle utilization are the two most important objectives of the framework. The authors in \cite{lin2018efficient} employ multi-agent RL, where each vehicle functions as an individual agent. Similarly, \cite{gueriau2018samod} provides a dynamic ride-sharing system in which both passenger assignments and fleet rebalancing are learned and performed by individual agents using multi-agent RL. Furthermore, in \cite{wen2017rebalancing}, the authors address rebalancing idle vehicles by developing a deep Q-learning approach.

There are well-known challenges associated with model-free RL algorithms. Examples of these difficulties include the necessity of a trade-off between policy exploitation and exploration, problem-dependent reward shaping, and the design of an appropriate neural architecture for the policy network. In addition, the majority of them are neither theoretically tractable nor can their convergence be investigated. Additionally, the performance of the controller may be significantly impacted by various formulations of the state and action spaces.

\subsubsection{Model-based}
Model-based AMoD techniques attribute an explicit model to system dynamics and utilize it to determine optimal decisions. Despite their complexity, they are powerful and allow us to examine the model's properties, including convergence. Numerous studies proposed and developed system models, including queuing \cite{zhang2016models,volkov2012markov}, fluidic \cite{pavone2012robotic}, network flow \cite{zhang2016control,rossi2018routing,carron2019scalable}, and data-driven \cite{lei2020efficient} approaches. Further classifications of model-based approaches include mathematical optimization and simulation-based methods. Various studies have tackled the rebalancing of vehicle fleets as a complex optimization problem. In \cite{war2019two}, dynamic repositioning and routing problems with stochastic demands are considered. The authors of \cite{zhang2018analysis} use queuing Jackson networks to propose a rebalancing algorithm. Reference \cite{chen2019d} approaches the problem as a mathematical program of static user equilibrium for a MoD system. Further, the solution is implemented as the basis for a linear program used to rebalance empty vehicles. Another way to model an AMoD system is to use a station-based arrangement such that each node in a graph represents a station that corresponds to an area in the city. The rebalancing can be done in such a setup on the graph nodes. A few other AMoD system models and theories have been developed in the literature, such as the linear optimization used in \cite{pavone2012robotic} to derive optimal rebalancing for the system at equilibrium. 
Moreover, a Markovian model has been proposed by \cite{volkov2012markov} to employ time-invariant control laws.

In network flow models, which are our interest, rebalancing problems are often formulated as (integer) optimization problems \cite{zgraggen2019model,carron2019scalable}, possibly considering competitive behavior \cite{martin2020rebalancing}.
Reference \cite{iglesias2018data} explores the stochastic models for future demand in conjunction with an MPC framework to study real-time rebalancing. \cite{dandl2019evaluating} enhances the result through machine learning. In order to address routing and rebalancing problems, it is common to study AMoD systems utilizing queuing network methods. Reference \cite{iglesias2019bcmp} offers novel tools to control and monitor the performance of AMoD systems and presents a queuing-theoretical framework to capture passenger arrivals, congestion impacts, vehicle routing, and vehicle battery charge levels. A theoretical framework is provided by \cite{wollenstein2020joint} for designing pricing and rebalancing strategies for AMoD systems using a fluid model with passenger and vehicle lines in each zone. Furthermore, a framework for optimization is developed to maximize profits while balancing the load on customers and vehicles in order to acquire the best rebalancing strategy given the endogenous travel demand rates.
\cite{rigas2018algorithms} addresses Electric Vehicle (EV) usage in Mobility-on-Demand (MoD) systems, aiming to maximize customer service and EV utilization. It introduces various algorithms, including a greedy heuristic and an incremental MIP, and evaluates their performance in different scenarios using real-world data.
\cite{drwal2017adaptive} introduces two pricing mechanisms for on-demand car rental systems to tackle unbalanced demand and reduce costly vehicle relocations. These mechanisms adjust prices based on station occupancy, customer valuations, and relocation costs. In \cite{wollenstein2021routing}, the authors provide a network flow model to enhance the routing and rebalancing tactics for intermodal AMoD systems, where autonomous cars collaborate with public transportation to deliver on-demand mobility services in mixed traffic (consisting of private and AMoD vehicles).
The primary goal is to enhance the service's quality by reducing total travel time while maintaining vehicle availability throughout the regions. It is demonstrated that pure AMoD systems can have a negative impact on system performance when there is a substantial amount of travel demand because rebalancing flows cause more traffic congestion. Additionally, they experimentally demonstrate that integrating AMoD systems with walking, micromobility, and public transportation may drastically cut down on overall travel times.

Combining the MPC algorithm with the network flow model offers an efficient tool for expressing complex constraints.
Additionally, MPC algorithms offer a powerful tool for exploiting time-varying travel forecasts.
The MPC algorithm is a well-suited approach to optimize vehicle rebalancing and scheduling in an AMoD system. In the MPC algorithm, an open-loop optimization problem is solved at each time step to produce a sequence of control actions up to a specific horizon and perform the first control action. Since MPC is an iterative-based algorithm, it can achieve closed-loop performance and robustness against model errors. It is also well-suited for constrained and complex systems. Furthermore, MPC’s predictive capabilities provide enhanced performance due to inherent model delays. Interested readers are referred to \cite{mayne2000constrained} for a comprehensive description of the MPC method. A few studies have applied the MPC to AMoD systems so far. In \cite{kang2021maximum}, the authors study an analysis of the max-pressure dispatch policy aimed to maximize customer throughput in the form of a MPC algorithm. In \cite{zhang2016model}, a MPC scheme is implemented for a constrained discrete-time model of an AMoD system. \cite{tsao2018stochastic} develops a stochastic MPC framework with an emphasis on vehicle dispatching and rebalancing allowing for the accounting of forecast uncertainty. Vehicles are initially distributed to passengers under the proposed controller, and later, empty vehicles are redistributed across the city. Furthermore, they demonstrate that the resulting integer linear optimization problem can be solved as linear programming thanks to the unimodularity of the problem. In \cite{alonso2017demand}, an integer linear program has been applied to reduce the dimensionality, assign vehicles to customers, and perform optimal rebalancing. Data-driven algorithms are proposed in \cite{miao2016taxi} and \cite{iglesias2018data}.
Reference \cite{iglesias2018data} develops an MPC algorithm leveraging historical data and neural networks to develop a model for short-term demand forecasts to address the dispatching and rebalancing problem. 
\cite{tsao2019model} proposes a real-time MPC framework to promote social welfare by optimizing the routes of both customer-carrying and empty vehicles, i.e., a weighted combination of vehicle mileage and passenger travel time. Moreover, a scalable MPC control has been developed in \cite{carron2019scalable} to keep the system balanced.

If the AMoD framework is not appropriately controlled, it can run into an imbalanced system, i.e., oversupplying the stations frequently opted as destinations. In contrast, regions with a high number of originating trips are undersupplied. To circumvent this issue, a rebalancing strategy is needed to move vehicles to stations in high demand. Aiming for the rebalancing strategy, we need a model to capture the dynamics of the AMoD system. The model proposed in this paper uses graph theory to capture the dynamics of the system and is well-suited for control purposes.

Users travel on paths that consist of multiple links. Therefore, demand is from origins to destinations as opposed to being associated with link demand. This is because links are the intermediate points on the paths that facilitate travel from origin to destination. In \cite{carron2019scalable}, the model is constructed based on the demand of each link on the paths. However, there is no such thing as link demand in transportation. Therefore, this model requires the estimation of the probability of the vehicles staying at the nodes or moving to the following links and the probability of entering the adjacent links (stopping and transition matrices). However, it is not an accurate representation of travel demand. As a result, we relax the model in \cite{carron2019scalable} by considering path demand instead of link demand. Although it can add computational complexity, it not only addresses the dependency of the links' demand, but we no longer need to estimate the stopping and transition matrices.

After proposing the model and an approximation for an AMoD system, a control framework needs to be implemented to ensure that the system is balanced.
Considering that the system is based on a linear model subject to a convex set of constraints, a natural approach is MPC. 
It satisfies system constraints and investigates the effectiveness of the model's predictive capabilities. MPC is also a robust algorithm due to its rolling-horizon structure. It requires recurrent state feedback to prevent the accumulation of errors in states over time.
In summary, the contributions of this paper can be expressed as follows:
\begin{enumerate}
\item Introduced a realistic and straightforward linear discrete-time dynamical model for the AMoD system.
\item Derived the desired rebalancing reference for the AMoD system model analytically, which will serve as the reference policy to find the optimal rebalancing of the AMoD system.

\item Implemented the MPC control framework to regulate the system to the optimal rebalancing and scheduling policy for the AMoD system.
\end{enumerate}

The remainder of the paper is organized as follows. In Section \ref{sec: system modeling} presents the problem formulation and model for the AMoD system, followed by some mathematical preliminaries. This section is concluded by discussing some properties of the approximated system. In Section \ref{sec: Implement MPC}, the MPC framework is presented. We present the results of the numerical case study examples in Section \ref{sec: results}. Finally, the paper is concluded in Section \ref{sec: conclusion}. We have summarized the notations and
abbreviations in Table \ref{tab: Notation}.

\begin{table*}[thbp]
    \centering
        \caption{Notation and abbreviations.}
{ \footnotesize   \begin{tabular}{c c}
    \hline   ${\bf 1}_m$ & Column vector of dimension $m$ with all entries equal to $1$.\\
    ${\bf 0}_n$ & Column vector of dimension $n$ with all entries equal to $0$.\\
    $\tilde {P}$ & Diagonal matrix with the elements of the vector ${\bf P} \in \mathbb{R}^n$ on the diagonal ($\tilde {P} = {\rm diag} \left ( {\bf P}\right) \in \mathbb{R}^{n\times n}$).\\
    $Q > 0\,(Q\geq 0)$ &  ${Q} \in \mathbb{R}^{n\times n}$ is (semi) positive definite.\\
${G}\left(N,A\right)$ & A digraph with a node set $N$ and link set $A$.\\
n,m & Cardinality of $N$ and $A$. \\
$E_{\rm in}$, $E_{\rm out}$, E& In-neighbors, out-neighbors, and incidence matrices ($E = E_{\rm in} -E_{\rm out}$).\\
t & Time step index.\\
$W_{rs}(t)$ & Number of customers willing to travel from station $r$ to station $s$.\\
$P_r(t)$ & Number of waiting or available vehicles at station $r$.\\
$F_{rs}(t)$& Number of vehicles moving along the link $\left\{ r,s \right\}$.\\
$V_{rs}(t)$& Number of customer-carrying vehicles travels from station $r$ to station $s$.\\
$U_{rs}(t)$& Dispatching rate from station $r$ to station $s$, which is necessary to maintain the queues bounded.\\
$R_{rs}(t)$& Rebalancing flow starting at station $r$ directed to station $s$.\\
$d_{rs}(t)$ & Arrival rate of customers in a time step by the realization of a Poisson process with parameter $\lambda_{rs}$.\\
$\lambda_{rs}$ & The parameter of a Poisson process.\\
$T_{rs}$ & Travel time from station $r$ to station $s$.\\
${\bf x}(t)$,${\bf v}(t)$ & The state and control at $t$.\\
$n_x$,$n_v$&Size of state and control variables. \\
$\bar{\bf x}$,$\bar{\bf v}$ & ${\bf x}(t)$,${\bf v}(t)$ in equilibrium.\\
${\mathcal A}, {\mathcal B},{\mathcal L}$& System dynamics.\\
QMPC$_{\rm QRef}$& Model predictive control with quadratic cost that tracks a reference obtained by quadratic programming.\\
QMPC$_{\rm LRef}$& Model predictive control with quadratic cost that tracks a reference obtained by linear programming.\\
LMPC$_{\rm QRef}$& Model predictive control with linear cost that tracking reference obtained by quadratic programming.\\
LMPC$_{\rm LRef}$& Model predictive control with linear cost that tracks a reference obtained by linear programming.\\
\hline
    \end{tabular}}
    \label{tab: Notation}
\end{table*}

\section{System Modeling}
\label{sec: system modeling}
This section explains how a directed graph can be transformed into a complete graph under a particular assumption. Also, a discrete-time linear dynamic model is formulated for the AMoD system. At the end of this section, some properties of the proposed model are discussed.
The symbols ${\bf 1}_m$ and ${\bf 0}_n$ denote column vectors of dimension $m$ and $n$ with all entries equal to $1$ and $0$, respectively. Given a vector ${\bf P} \in \mathbb{R}^n$, we define $\tilde {P} = {\rm diag} \left ( {\bf P}\right) \in \mathbb{R}^{n\times n}$ as a diagonal matrix with the elements of the vector $\bf p$ on the diagonal.
\subsection{Preliminaries}
Consider a directed graph ${G}\left(N,A\right)$ where $N=\left\{1, \dots, n\right\}$ is the set of nodes and $A=\left\{1, \dots, m\right\}$ is the set of links. Let $E_{\rm in}$ and $E_{\rm out}\in \left\{0,1\right\}^{n\times m}$ be the in-neighbors and out-neighbors matrices.
If a link enters a node, the associated entry is one in $E_{\rm in}$. Similarly, if a link exits a node, the associated entry equals one in the matrix $E_{\rm out}$. The incidence matrix $E \in \left\{ -1,0,1\right\}^{n \times m}$ can be derived by $E = E_{\rm in} -E_{\rm out}$.
\subsection{Model}
We relax the model in \cite{carron2019scalable} by considering path demand instead of link demand. We transform the graph into a complete graph of origins and destinations (point-to-point travel) rather than a road network. A complete graph is a graph in that each node is connected to the other nodes, either with an actual link for adjacent nodes or a virtual link for non-adjacent nodes. If there is no actual link between two nodes, we consider the shortest path between those nodes as a virtual link. In this network transformation, we no longer need to use and estimate the stopping and transition matrices like in \cite{carron2019scalable}. These features result in a realistic and straightforward linear discrete-time model for an AMoD system.
\begin{figure}[thbp]
  \centering
  \begin{subfigure}{0.45\textwidth}
    \centering
  \includegraphics[width=0.9\textwidth]{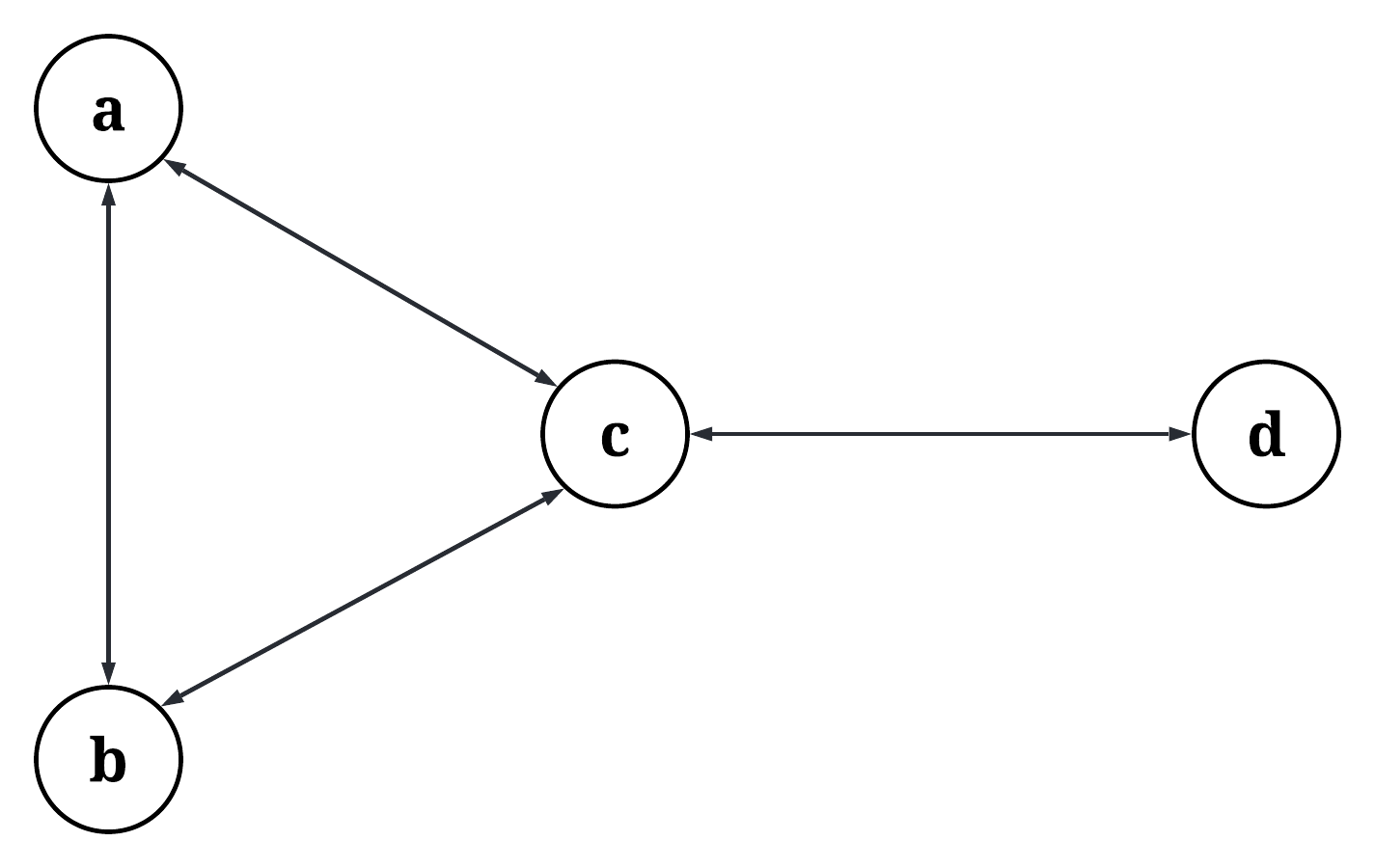}
  \caption{ }
  \label{fig:RealNetwork}
  \end{subfigure}
\begin{subfigure}{0.45\textwidth}
    \centering
  \includegraphics[width=0.9\textwidth]{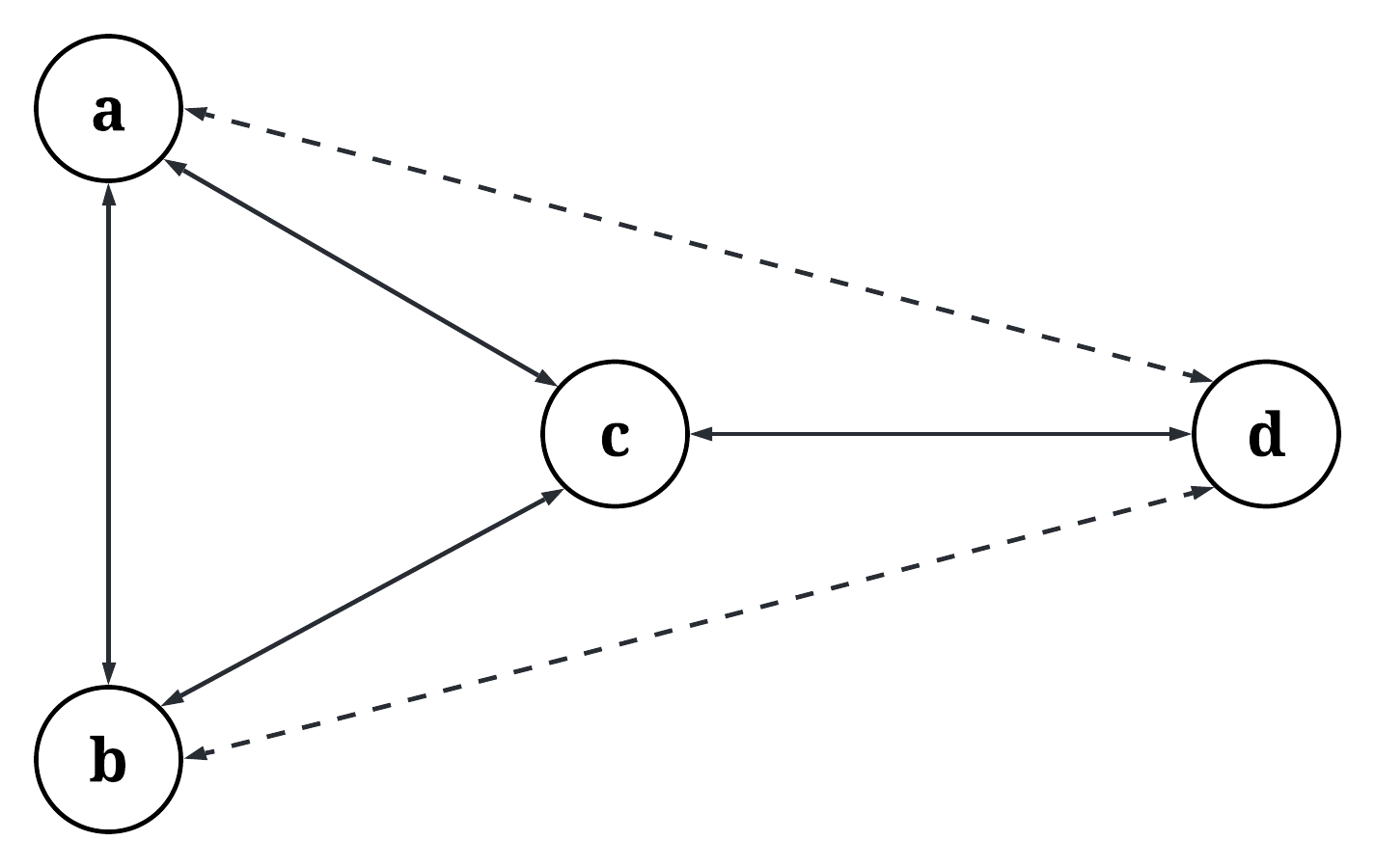}
 \caption{ }
  \label{fig:TransformedNetwork}
  \end{subfigure} 
  	\caption{A network (graph) with four stations. (a) The original network; (b) The transformed (complete) network.}
  	  \label{fig:WNetwork}
  	\end{figure}

Figure \ref{fig:RealNetwork} and \ref{fig:TransformedNetwork} show a network (graph) and its transformation (complete graph), respectively. Note that solid lines and dashed lines in figure \ref{fig:WNetwork} denote the actual and virtual links, respectively.
\begin{Assumption}
\label{assumption1}
Graph $G$ is strongly connected, i.e., there is at least one path between every two nodes.  
\end{Assumption}
Assumption \ref{assumption1} is crucial for transforming the graph $G$ to a complete graph $\hat G$. In complete graph $\hat G$, the number of the nodes does not change compared to the directed graph $G$; however, the number of the links is $\frac{n(n-1)}{2}$.

The proposed linear discrete-time time-delay dynamic system is as follows:
\begin{subequations}
\begin{align}
\label{LinearModel_a}
W_{rs}\left ( t+1 \right )= &W_{rs}\left ( t \right ) +d_{rs}\left ( t \right )-V_{rs}\left ( t \right ),
\\
\label{LinearModel_b}
P_{r}\left ( t+1 \right )=  &P_{r}\left ( t\right )-\sum_{s\in N} {\left( V_{rs}\left ( t \right ) 
+R_{rs}\left ( t \right )\right ) }+\sum_{q\in N} {\left( U_{qr}\left ( t-T_{qr} \right )
+R_{qr}\left ( t - T_{qr}\right )\right ) },\\
\label{LinearModel_c}
F_{rs}\left ( t+1 \right )=&F_{rs}\left ( t\right )+ V_{rs}\left ( t \right )+R_{rs}\left ( t \right )- V_{rs}\left ( t - T_{rs}\right ) - R_{rs}\left ( t - T_{rs} \right ),
\end{align}
\label{LinearModel}
\end{subequations}

${\forall \, r , s \in N}$ where state variable $W_{rs}$ denotes the waiting customers along link $\left \{ r,s \right \}$, i.e., customers willing to travel from station $r$ to station $s$. The term $d_{rs}(t)$ represents the arrival of customers in a time step given by the realization of a Poisson process of parameter $\lambda_{rs}$.
It should be noted that the passengers are assumed to always stay in the system, as implied by \eqref{LinearModel_a}.
State variable $P_r$ characterizes the waiting or available vehicles at station $r$.
State variable $F_{rs}$ denotes vehicles moving along the link $\left\{ r,s \right\}$, including both customer-carrying and rebalancing vehicles. Control input $V_{rs}$ is the number of available vehicles at station $r$ with a customer that will be dispatched to link $\left\{ r,s \right\}$. $V_{rs}\left ( t \right ) = \min (W_{rs}\left ( t \right ), U_{rs}\left ( t \right ))$, where $U_{rs}(t) \geq \lambda_{rs}$, $\forall r,s \in N^2$.
This is a necessary condition for the total number of customers at each station is remain bounded and we will assume it throughout the paper.
$R_{rs}$ is the number of available vehicles at station $r$ that will be dispatched to link $\left\{ r,s \right\}$ for rebalancing.
Equation \eqref{LinearModel_a} describes the evolution of the passengers at $r$ headed to $s$, equation \eqref{LinearModel_b} captures the evolution of the available vehicle, and equation \eqref{LinearModel_c} describes the evolution of link flow dynamics for customer-carrying and rebalancing vehicles.
Note that each vehicle serves only one customer request at a time, i.e., sharing/pooling is not considered.
 
We assume that the travel times $T_{rs}$ are considered constant and exogenous. However, they may not be necessarily the same value at different times of the day, i.e., $T_{rs}^t$ instead of $T_{rs}$.
The reason for assuming constant travel time is that the number of AMoD vehicles is much less than the rest of the traffic. However, an exogenous {\it uncertainty} will be added to the travel times in the simulation to evaluate the robustness of the system and the control framework.
\begin{Remark}
Passenger demand is considered according to some scenarios. In each scenario, we consider that the passenger demand follows a Poisson distribution with a constant rate. Considering different scenarios is essential to investigate how well the proposed methodology and method work in such scenarios, especially in the morning and noon, when demand is higher than in the afternoon.
\end{Remark}

Delays in model \eqref{LinearModel} pose a challenge in deriving the control law. In order to avoid delays in the model, an approximation used in \cite{carron2019scalable} is implemented. By considering a first-order lag approximation of the time delays, it is assumed that the number of vehicles exiting a link is proportional to the number of vehicles on that link. In other word, at each time instant $t$, the quantity $F_{rs}\left ( t \right )/T_{rs}$ leaves the link $\left \{ r,s \right \}$. Therefore, 
$V_{rs}\left ( t-T_{rs} \right ) +R_{rs}\left ( t - T_{rs}\right )$ can be replaced by $F_{rs}\left ( t \right )/T_{rs}$. Consequently, linear model \eqref{LinearModel} can be rewritten as follow:
\begin{subequations}
  \begin{align}
  \label{Apr_LinearModel_a}
W_{rs}\left ( t+1 \right )= &W_{rs}\left ( t \right ) +d_{rs}\left ( t \right )-V_{rs}\left ( t \right ),\\
  \label{Apr_LinearModel_b}
P_{r}\left ( t+1 \right )=  &P_{r}\left ( t\right )-\sum_{s\in N} {\left( V_{rs}\left ( t \right ) 
+R_{rs}\left ( t \right )\right ) }+\sum_{q\in N} {\left(\frac{F_{qr}\left ( t\right )}{T_{qr}} 
\right )},\\
  \label{Apr_LinearModel_c}
F_{rs}\left ( t+1 \right )=&\left ( 1- \frac{1}{T_{rs}}\right )F_{rs}\left ( t\right )+ V_{rs}\left ( t \right )+R_{rs}\left ( t \right ),
\end{align}
\label{Apr_LinearModel}
\end{subequations}
${\,\forall \, r , s \in N}$. This AMoD system is subject to some constraints itemized as follows:
\begin{enumerate}
\item The number of customer-carrying vehicles traveling from station $r$ to station $s$ is limited by the current queue length and the dispatching rate from station $r$ to station $s$, i.e., 
$$V_{rs}( t ) = \min (W_{rs}( t ), U_{rs}(t)).$$
\item  
Given the current availability, the number
of vehicles dispatched at any time step is limited.
\item  
Given the current availability, the number
of vehicles dispatched at any time step is limited.
\begin{equation}
    \sum_{s\in N}{V_{rs}(t)+R_{rs}(t)}\leq P_r(t).
\end{equation}
\item  These constraints enforce the non-negativity of state and control input variables.
\begin{align}
P_{r}\left ( t \right )\geq 0,\quad
V_{rs}\left ( t \right )\geq 0,\quad
R_{rs}\left ( t \right )\geq 0.
\end{align}
\end{enumerate} 
Note that constraints are valid for every $r$ and $s$ in $N$.

The global system associated with graph $\hat G$ is represented as
  \begin{equation}
  {\bf x}\left ( t+1 \right )={\mathcal A}{\bf x}\left ( t \right )+{\mathcal B}{\bf v}\left ( t \right )+{\mathcal L}{\bf {d}}\left ( t \right ),
       \label{SystemVector}
   \end{equation}
where the vector of all state variables ${\bf x}\in {\mathbb{R}^{2n^2-n}}$ is $\left ({\bf W}\left ( t\right ),{\bf P}\left ( t\right )
,{\bf F}\left ( t\right )\right )^T$ and the vector of all control input variable ${\bf v}\left ( t\right ) \in \mathbb{R}^{{2n(n-1)}}$ is define as ${\bf v}\left ( t\right )=\left ({\bf V}\left ( t\right ),{\bf R}\left ( t\right )
\right )$. ${{\bf d}(t)} \in {\mathbb{R}^{n(n-1)}}$ represents arriving customers. Matrices ${\mathcal A}$, ${\mathcal B}$, and can be written as below:
  \begin{align}
\mathcal A= 
\begin{bmatrix}
 I_{n(n-1)} & 0 & 0\\ 
 0 & I_n & E_{\rm in}\tilde{T}^{-1}\\ 
 0 & 0 & I_{n(n-1)}-\tilde{T}^{-1}
\end{bmatrix}, \quad \mathcal B=
\begin{bmatrix}
-I_{n(n-1)} & 0\\ 
-E_{\rm out}& -E_{\rm out}\\ 
I_{n(n-1)} & I_{n(n-1)}
\end{bmatrix}, \quad \mathcal{L} = \begin{bmatrix}
I_{n(n-1)}\\ 
0\\
0
\end{bmatrix}.
 \end{align}

\subsection{Model Properties}
In this section, some properties of the model are investigated. First, the well-posedness of the model is proven. Then, a set of equilibria is derived. By solving an optimization problem, the desired rebalancing number can be derived.

In the following proposition, the well-posedness of the proposed model \eqref{Apr_LinearModel} is validated.
\begin{Proposition} 
\label{well-posedness}
If graph $G$
is strongly connected, then
  \begin{equation}
  {\bf 1}_{n}^{T}{\bf P}\left ( t \right )
  +{\bf 1}_{n(n-1)}^{T}{\bf F}\left ( t \right )
  =M \quad \forall t \geq 0.
  \end{equation}
In other words, the number of vehicles in the network at any time is constant.
\end{Proposition}
\begin{proof}
We prove the well-posedness by showing the conservation of the number of vehicles at each time step. The total number of cars at
time instant $t + 1$, i.e., $M\left( t+1\right )$ is as follow:
   \begin{flalign}
  M\left( t+1\right )
  =&{\bf 1}_{n}^{T}{\bf P}\left ( t+1 \right )
  +{\bf 1}_{n(n-1)}^{T}{\bf F}\left ( t+1 \right )  \notag \\
  =&{\bf 1}_{n}^{T}{\bf P}\left ( t \right )-{\bf 1}_{n}^{T}E_{\rm out}\left ({{\bf V}}\left ( t \right )+{{\bf R}}\left ( t \right )  \right )
  +{\bf 1}_{n}^{T}E_{\rm in}\tilde{T}^{-1}{\bf F}\left ( t \right )  \notag\\
  & +{\bf 1}_{n(n-1)}^{T}\left (    \left ( \mathbb{I}-\tilde{T}^{-1} \right ){\bf F}\left ( t \right ) \right )+   {\bf 1}_{n(n-1)}^{T}\left ({{\bf V}}\left ( t \right )\!+\!{{\bf R}}\left ( t \right )  \right ).
 \end{flalign}
By using the equalities ${\bf 1}_{n}^{T}E_{\rm in}={\bf 1}_{n}^{T}E_{\rm out}={\bf 1}_{n(n-1)}^{T}$, we have 
   \begin{flalign}
  M\left( t+1\right )
  =&{\bf 1}_{n}^{T}{\bf P}\left ( t \right )-{\bf 1}_{n(n-1)}^{T}\left ({{\bf V}}\left ( t \right )+{{\bf R}}\left ( t \right )  \right )
  +{\bf 1}_{n(n-1)}^{T}\tilde{T}^{-1}{\bf F}\left ( t \right )  \notag\\
  &+
  {\bf 1}_{n(n-1)}^{T}\left (    \left ( \mathbb{I}-\tilde{T}^{-1} \right ){\bf F}\left ( t \right ) \right )+   {\bf 1}_{n(n-1)}^{T}\left ({{\bf V}}\left ( t \right )+{{\bf R}}\left ( t \right )  \right )  \notag\\
  =&{\bf 1}_{n}^{T}{\bf P}\left ( t \right )
  +{\bf 1}_{n(n-1)}^{T}{\bf F}\left ( t \right )  = M\left( t\right ).
  \end{flalign}
It is showing that for $\forall t \geq 0$, $M\left( t+1\right )=M\left( t\right )=M$.
\end{proof}
\begin{Proposition}[Equilibrium Points]
\label{Proposition_Equilibrium}If graph $G$ is strongly
connected and $d_{rs}$ = $\lambda_{rs}$ for $\forall \left \{r, s\right \} \in {\hat A}$, where $\lambda_{rs}$ represents the Poisson arrival rate for the link $\left \{r, s\right \}$, then equilibrium points of system \eqref{SystemVector} are given by 
${\bar{\bf x}}=\left ({\bar{\bf W}},{\bar{\bf P}}
,{\bar{\bf F}} \right )$, where ${\bar{\bf W}}$ and ${\bar{\bf P}}$
can be any arbitrary positive vector, ${\bar{\bf F}}=\tilde{T} \left (  {\bar{\bf \lambda}} + {\bar{\bf R}}\right )$, ${\bar{\bf V}}={{\bf \lambda}}$, and ${\bar{\bf R}}$ satisfies 
$\left ( E_{\rm in}-E_{\rm out}\right )\left ({\bar{\bf R}}+{{\bf \lambda}}\right )=E\left ({\bar{\bf R}}+{{\bf \lambda}}\right )=0$.
\end{Proposition}
\begin{proof}
${\bf x } \left(t+1\right)={\bf x } \left(t\right)$ for $\forall t \in \mathbb{N}$ holds for a linear discrete-time system at the equilibrium.
By using this equation and also model \eqref{Apr_LinearModel}, equilibrium points can be derived.
\end{proof}
\begin{Corollary}
If the number of nodes, $n$, is greater than 2, there will be an infinite number of equilibrium points. Also, the desired equilibrium point that minimizes the number of rebalancing can be found by solving an optimization problem.
\end{Corollary}
\begin{proof}
Equation $E\left ({\bar{\bf R}}+{{\bf \lambda}}\right ) = 0$ yields a system of $n$ equations in $n(n-1)$ unknowns, and if $n>2$, the number of solutions is infinite. Thus, we are able to determine the desired equilibrium point that minimizes the number of rebalancing vehicles required weighted by the associated travel time as the following optimization problem:
\begin{subequations}
\begin{flalign}
\label{rebalnce_optimization_cost}
\mathop {\min {\mkern 1mu} }\limits_{\bf R}\quad& J({\bf R})\\
\rm{s.t.}\quad&E\left ({{\bf R}}+{{\bf\lambda}}\right ) = 0\\
&{{\bf R}}\geq 0.
\end{flalign}
\label{OptForRebalance}
\end{subequations}
The cost function $J({\bf R})$ can be defined either as a linear penalty
function
$J({\bf R}) = \left\| {\tilde T }{\bf R} \right\|_1$
or quadratic 
$J({\bf R}) = \left\| {\tilde T }^{\frac{1}{2}}{\bf R} \right\|_2^2$.
Note that since travel times are positive, ${\tilde T }$ is full column rank.
Unlike the linear cost, a quadratic cost aims to reduce large deviations from the equilibrium.

\end{proof}
\begin{Corollary}
When $\lambda$ satisfies $E\lambda = 0$ in \eqref{OptForRebalance}, rebalancing is not required.
\end{Corollary}
As a special case of $E\lambda = 0$, $\lambda^{ij}=\lambda^{jk}$ for $\forall i,j,k$, is a sufficient condition implying rebalancing is not required. However, this is a solid condition that is very unlikely to happen in a real network.

\section{MPC Implementation}
\label{sec: Implement MPC}
\begin{figure*}[th]
  \centering
 \includegraphics[width=1\textwidth]{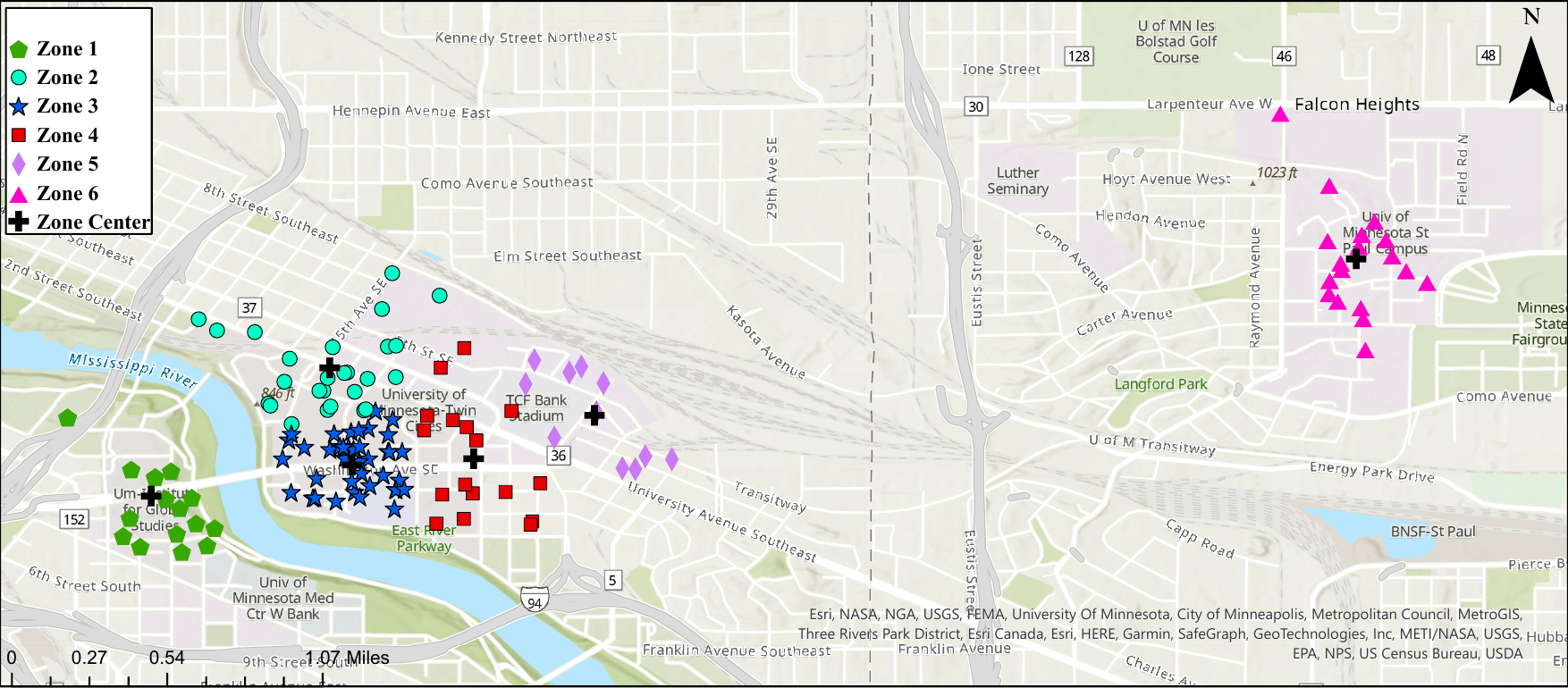}
  \caption{The University of Minnesota-Twin Cities campus network map for testing the proposed model and control strategy. Zone 1 denotes the west bank of the Minneapolis campus, Zones 2, 3, 4, and 5 represent the east bank of the Minneapolis campus, and Zone 6 denotes the St. Paul campus. Zones have been extracted using Lloyd’s k-mean algorithm and travel data.}
  \label{fig:Campus Network}
\end{figure*}
This section presents the MPC algorithm for regulating the AMoD system around the desired equilibrium points.
Also, at the end of this section, the scalability and computational complexity of the MPC approach are discussed.
\subsection{MPC Implementation}
The MPC framework for regulation around the equilibrium point
$\bar {\bf x}= \left ({\bar {\bf W}},{\bar {\bf P}},{\bar{\bf F}}\right )^T$ and
$\bar {\bf v}= \left ({\bar {\bf V}},{\bar {\bf R}}\right )^T$
is expressed by the finite-horizon control problem and as follows:
\begin{subequations}
\begin{align}
\label{maincost}
\min_{\Delta v}& \sum_{i=0}^{N_{\rm {hor}}-1}{J(\Delta{\bf x}_{t+i},\Delta{\bf v}_{t+i})}
\\
{\rm s.t.}&\notag\\
&\Delta {\bf x}_{t+i+1}=\mathcal A\Delta {\bf x}_{t+i}+\mathcal B\Delta{\bf v}_{t+i}
\\
&\left ( {\bar {\bf x}}+\Delta {\bf x}_{t+i},{\bar {\bf v}}+\Delta {\bf v}_{t+i} \right )\in \mathcal{X}\times\mathcal{V}
\\
& \Delta {\bf x}_{t+N_{\rm {hor}}}=0
\\& i=\left \{0,\dots,N_{\rm {hor}}-1 \right\},
\end{align}
\label{Main_optimization}
\end{subequations}
where $\Delta {\bf x}\left(t\right)={\bf x} - {\bar{\bf x}}$ and $\Delta {\bf v}\left(t\right)={\bf v} - {\bar{\bf v}}$. $N_{\rm {hor}}$ is the optimization or control horizon. 
The objective function $J({\bf R})$ can be defined either as a linear penalty function
${J(\Delta{\bf x}_{t+i},\Delta{\bf v}_{t+i})} = \left\| 
{Q\Delta {\bf x}_{t+i}}
\right\|_1 +\left\| 
{S\Delta {\bf v}_{t+i}}
\right\|_1$ or quadratic 
${J(\Delta{\bf x}_{t+i},\Delta{\bf v}_{t+i})} = \left\| 
{Q^{\frac{1}{2}}\Delta {\bf x}_{t+i}}
\right\|_2^2 + \left\| 
{S^{\frac{1}{2}}\Delta {\bf v}_{t+i}}
\right\|_2^2$. $Q$ and $S$ are positive semidefinite matrices. By choosing weights matrices $Q$ and $S$ as
 \begin{equation}
 Q= \begin{bmatrix}
{\tilde {\bf\lambda}} & 0 & 0\\ 
0 & 0 & 0\\ 
0 & 0 & 0
\end{bmatrix}
\quad \quad
S =\begin{bmatrix}
0 & 0 \\ 
0 & {\tilde T}
\end{bmatrix},
\label{eq:weights}
\end{equation}
we are emphasizing the high-demand routes as well as longer paths compared to the other paths. Since we wish to choose minimum equilibrium for $\bf w$, i.e., ${\bf w } = {\bf 0}_{n^2-n} $, the first term in the quadratic case of cost function \eqref{maincost} is equal to ${\bf W}_{t+i}^{T} {\tilde {\bf\lambda}} {\bf W}_{t+i}$. It implies that deviations from the equilibrium point are minimized by service quality reduction for customers waiting at stations with short queues. The optimal solution of the optimization Problem \eqref{Main_optimization} is a feedback of the state variables, i.e., 
${\bf v}^{*}\left ( \Delta {\bf x} \right ) = \begin{Bmatrix}
{\bf v}_{0}^{*}\left ( \Delta {\bf x} \right ),\dots, {\bf v}_{N_{\rm {hor}}-1}^{*}\left ( \Delta {\bf x} \right )
\end{Bmatrix}$.

Intuitively, when the exogenous disturbance ${\bf d}(t)$ is zero for $t > 0$ (no new customers arrive after time zero), by selecting feasible inputs $V$, ${\bf W} \to 0 $ as $t \to {\infty}$.
\begin{Theorem}[Closed-Loop Stability]
Let $D = \{{\bf d}: \forall {\bf U}(t) \in {\mathcal U} \text{ s.t. } {\bf U}(t) \geq \lambda \}$ and $D^{0}$ be the interior of $D$. If ${\bf d} \in D^0$, then the dispatching policy obtained by the MPC framework is stabilizing.
\label{theorem:MPC_convergence}
\end{Theorem}
\begin{proof}
The proof of Theorem \ref{theorem:MPC_convergence} can be performed by implementing the Lyapunov Drift Theorem (\cite{neely2022stochastic}). Further details of the proof can be found in Appendix \ref{Theorem: Proof of Main Theorem}.
\end{proof}
\begin{Remark}
MPC input is as $u_{MPC} = u_I + u_F$, where $u_I$ and $u_F$ are the integers and the fractional parts, respectively.
By drawing a sample $q$ from a uniform distribution ${\mathcal U}(0, 1)$, if $u_F > q$, we apply the input $u = u_I + 1$; otherwise, $u = u_I$. It is crucial to verify that no constraint is violated.
\end{Remark}
\subsection{Scalability and Computational Complexity} 
For large-scale MPC problems, e.g., $n_x > 10$, online solution methods are used. The two most prominent methods are active set and interior point methods.
The computational complexity of interior point methods is $\mathcal{O}(N_{\rm hor}(n_x + n_u)^3)$ (\cite{wang2009fast,frison2013fast,richter2011computational}).
Since we consider the control horizon $N_{\rm hor}$ in any discussed networks to be constant, the only concern is the size of the state variables $n_x = {2n^2-n}$ and control inputs $n_v = {2n^2-2n}$. Hence, the order of computational complexity is $\approx \mathcal{O}(n^6)$. Although computational complexity can grow rapidly when the number of zones $n$ increases, it is still polynomial. Apart from that, in order to improve/reduce the computational complexity, proper partitioning of a large network into a reasonable number of zones based on origin-destination demands could be done, e.g., using appropriate clustering algorithms such as Lloyd’s k-Means algorithms.

\section{Simulation Studies}
\label{sec: results}
In this section, first, we introduce a network for the tests we perform.
Second, we present another state-of-the-art rebalancing algorithm to be compared with the MPC framework. Finally, two case studies are demonstrated to examine the properties of the MPC.
\subsection{Studied Network}
The University of Minnesota-Twin Cities (UMN) campus network shown in Fig. \ref{fig:Campus Network} is considered a site on which to perform the tests in some realistic scenarios. There are 132 spots (buildings) demanded by the customers.

UMN's campuses are partitioned into $n$ virtual zones, in order to obtain the associated digraph.
Choosing a high number of zones has a few advantages, such as high network coverage, high modeling accuracy, and high accessibility for passengers. However, choosing a high number of zones increases the computational time, as it was shown that the order of computational complexity is $\mathcal{O}(N_{\rm hor} n^6)$. Therefore, there should be a trade-off between modeling accuracy and computational efficiency. Based on these factors and the origin-destination demands extracted from the historical data and GIS analysis, we chose $n=6$ that reasonably covers all the campuses and has an admissible computational time.
The partitioning can be obtained by various techniques where we rely on Lloyd’s k-mean algorithm \cite{lloyd1982least}.
It should be noted that the rebalancing performance is certainly affected by partitioning, but a detailed analysis is beyond the scope of this article.

\begin{figure}[h]
  \centering
 \includegraphics[width=.6
 \textwidth]{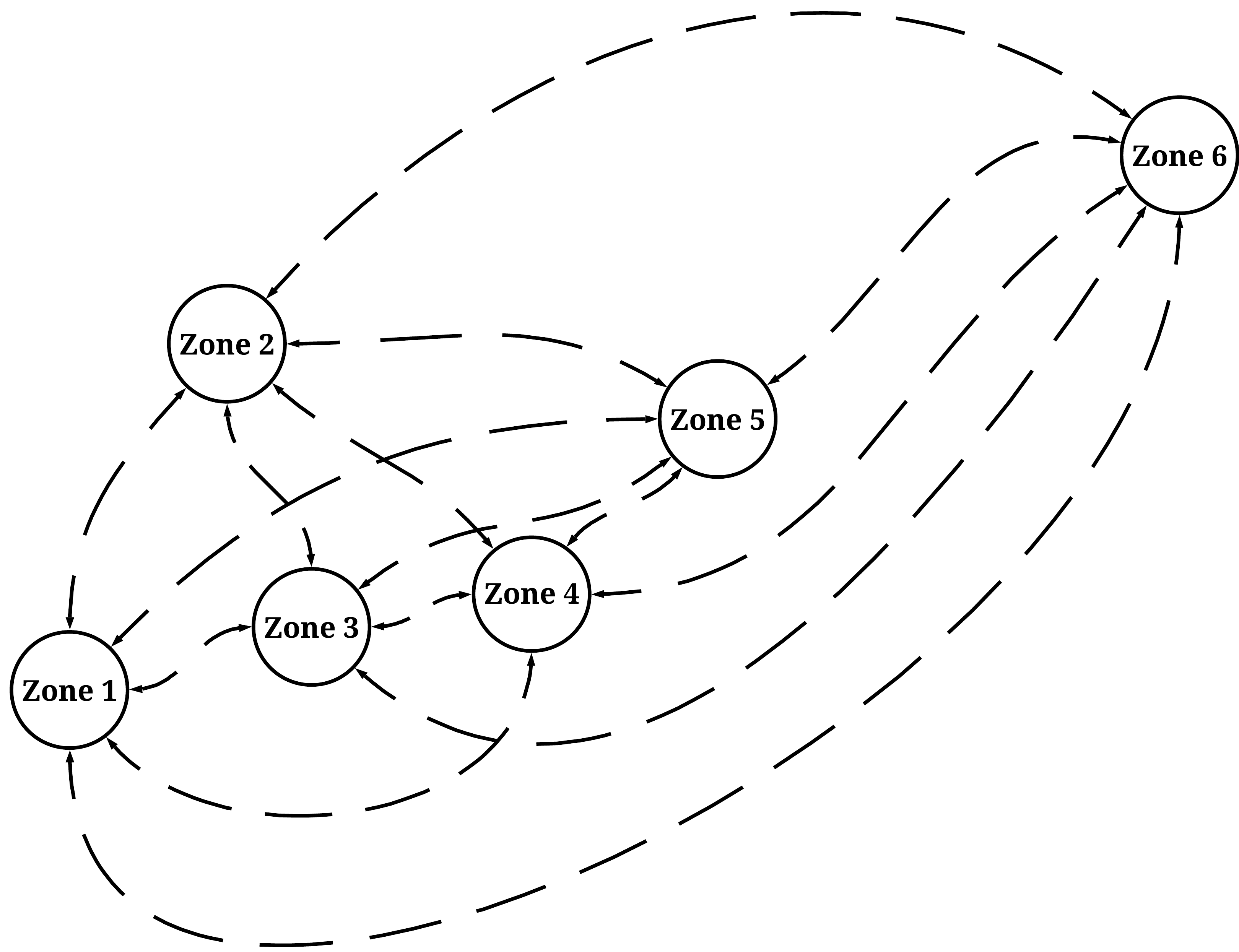}
  \caption{The University of Minnesota zones.}
  \label{fig:Campus zones}
\end{figure}

AMoD is considered the unique provider of mobility services.
A digraph with $n = 6$ vertices and $m = 30$ links is produced by partitioning; the graph vertices are superimposed on the map shown in Fig. \ref{fig:Campus zones}.

\begin{figure}[h]
  \centering
 \includegraphics[width=0.6\textwidth]{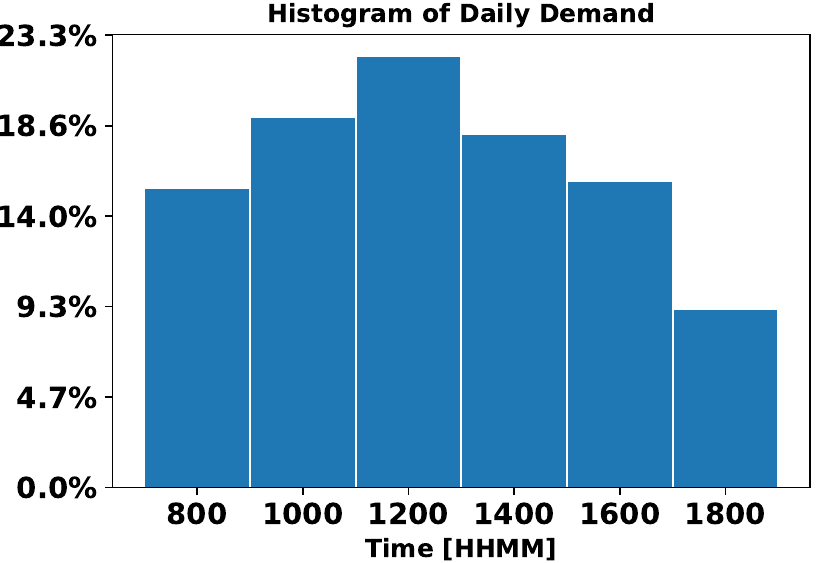}
  \caption{Histogram of the daily demand for the UMN's campuses.}
  \label{fig:demand_histogram}
\end{figure}

Figure \ref{fig:demand_histogram} shows the histogram of the daily demand for the UMN’s campuses. The peak occurs between 11:00 AM and 1:00 PM.
Since the demand represents the intra-zonal trips on the campuses (not commutes to the campus), it does not necessarily follow the typical morning and afternoon peaks.


\subsection{Improved adaptive real-time rebalancing (IARR)}
To further investigate the performance of the MPC controller, a comparison is carried out with a state-of-the-art rebalancing algorithm, namely adaptive real-time rebalancing (ARR), \cite{pavone2012robotic}. In this approach, the optimal rebalancing policy can be found as the solution to a linear program. This algorithm minimizes the average number of empty (rebalancing) vehicles traveling in the network. 
We further modified the real-time rebalancing and included the dispatching rate in the objective function weighted by the average arrival rate ($\lambda$).
At each time step, using the current state 
${\bf x}(t) = \begin{bmatrix}
{\bf W}( t ),{\bf P}( t )
,{\bf F}(t)\end{bmatrix}^T$, and following the optimization problem, the desired rebalancing and dispatching policies will be derived.

\begin{subequations}
\begin{align}
\mathop {\min {\mkern 1mu} }\limits_{{\bf R}(t),{{\bf U}(t)}}\quad & {\bf T}^T{\bf R}(t) - {\bf \lambda}^T{\bf U}(t)\\
{\rm s.t.}\quad & E_{\rm out}({\bf V}(t) + {\bf R}(t)) \leq {\bf P}(t)+  E_{\rm in}\tilde{T}^{-1} {\bf F}(t)\\
\quad & {\bf U} (t) \geq \lambda\\
\quad &{\bf R} (t) \geq 0.
\label{eq: IARR}
\end{align}
\label{eq: IARR}
\end{subequations}
Note that $V(t) = \min ({\bf W}(t),{\bf U}(t))$.
IARR aims to minimize the number of required rebalancing actions while maximizing service rates, which, in turn, reduces queue lengths and enhances service quality.

\subsection{Case Studies}
Two case studies are considered to evaluate the performance of the MPC. Comparisons are performed by applying MPC and ARR. In Case Study 1,
we consider each time step equal to two minutes, while in Case Study 2,
we consider each time step equal to three minutes. It means dispatching customers and rebalancing are performed every two minutes and three minutes in Case Studies 1 and 2. 
The control horizon is $N_{\rm hor} = 8$.

The UMN historical data are processed over 12 hours.
Many pairs are more frequently used than others, which implies a significant imbalance in the origin-destination pairs.
The fleet size is 125, which serves 2,940 mobility requests.
The reference being tracked with the MPC controller is recomputed via Problem \eqref{OptForRebalance} every 2 hours implementing a quadratic and linear cost. Afterward, the MPC algorithm is tested with four different scenarios: QMPC$_{\rm QRef}$, QMPC$_{\rm LRef}$, LMPC$_{\rm QRef}$, and LMPC$_{\rm LRef}$. QMPC denotes the MPC if cost \eqref{maincost} is quadratic, while LMPC represents the MPC if cost \eqref{maincost} is linear, no matter the type of reference, linear or quadratic.
The subscript ${\rm QRef}$ and ${\rm LRef}$ indicate whether the cost function in \eqref{rebalnce_optimization_cost} is quadratic or linear, respectively.


In Proposition \ref{well-posedness}, we showed that the number of vehicles is constant at each time step (including equilibrium) and is equal to ${\bf 1}_{n}^{T}{\bf P}\left ( t \right ) +{\bf 1}_{n(n-1)}^{T}{\bf F}\left ( t \right )$. Therefore, $\underline{M}= {\bf 1}_{n(n-1)}^{T}{\bar{\bf F}}={\bf T} \left (  {\bar{\bf \lambda}} + {\bar{\bf R}}\right )$ can be considered a lower band for the fleet size.
We used Dijkstra’s algorithm \cite{dijkstra1959note} to compute the shortest path between the zone centers on a real road network (Google Maps).
Initial conditions for the AMoD model are $\bf{x}(0)=\begin{bmatrix}
0^T_{\frac{n(n-1)}{2}} & \frac{\underline{M}}{n}{1^T_{n}} & 0^T_{\frac{n(n-1)}{2}}  
\end{bmatrix}^T$.
The average queue length, average waiting time,
and the average empty distance, i.e., the number of miles driven by cars without customers, are the metrics that we use to compare the algorithms.
\begin{Remark}
In all the simulations performed, no congestion effects have been considered, i.e., travel times are considered exogenous. If congestion is considered in the model, travel times are endogenous and a function of the policies performed by Problem \eqref{Main_optimization}. In that case, the model is no longer linear and a detailed analysis is beyond the scope of this article.
\end{Remark}

\begin{figure*}[thbp]
\begin{subfigure}{0.48\textwidth}
    \centering
  \includegraphics[width= 1\textwidth]{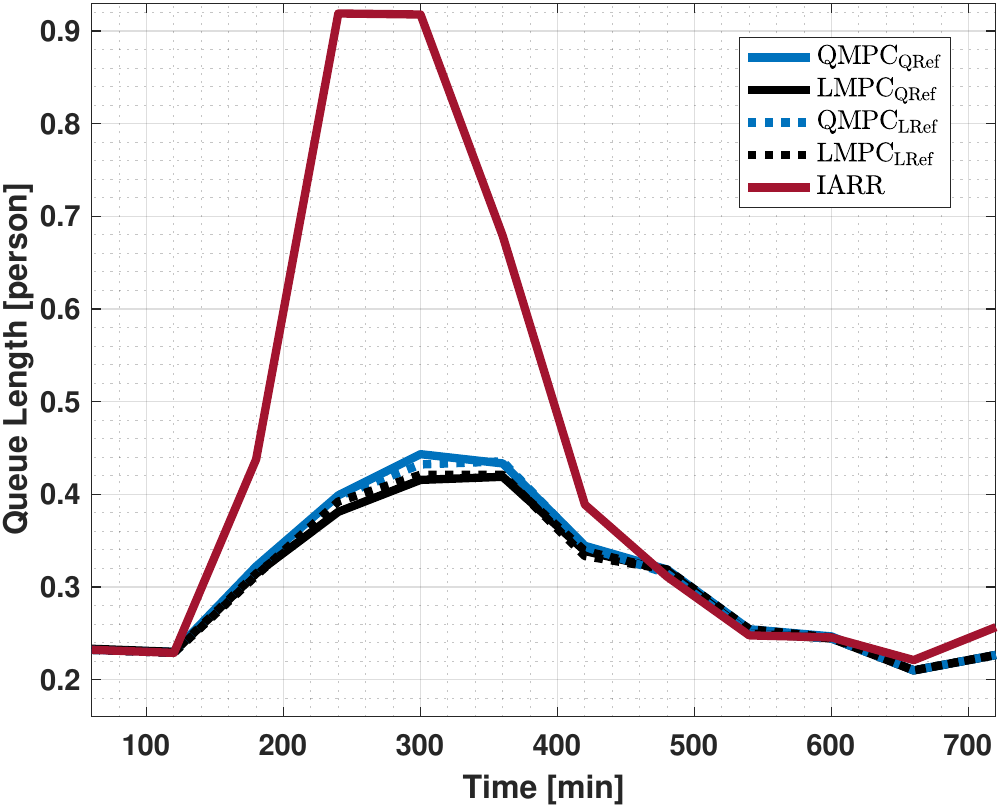}
 \caption{Case Study 1: The average queue length.}
   \label{fig: case1 W}
  \end{subfigure}
  \hfill
\begin{subfigure}{0.48\textwidth}
    \centering
  \includegraphics[width=1\textwidth]{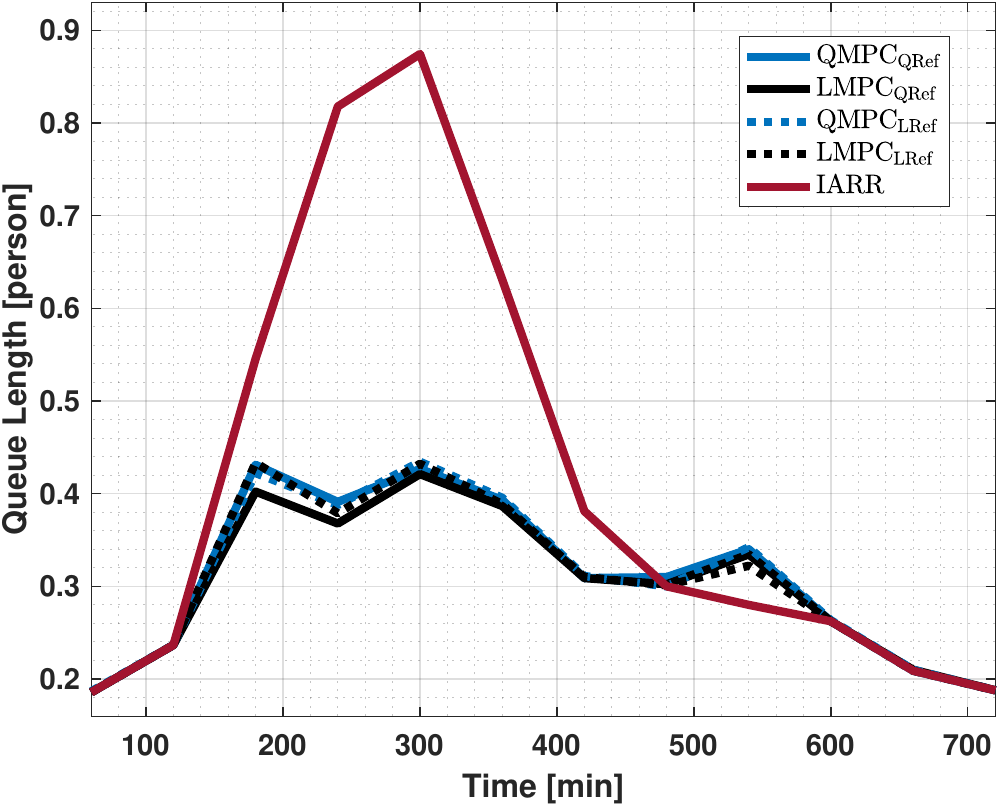}
 \caption{Case Study 2: The average queue length.}
   \label{fig: case2 W}
  \end{subfigure}
\hfill
\begin{subfigure}{0.48\textwidth}
    \centering
  \includegraphics[width=1\textwidth]{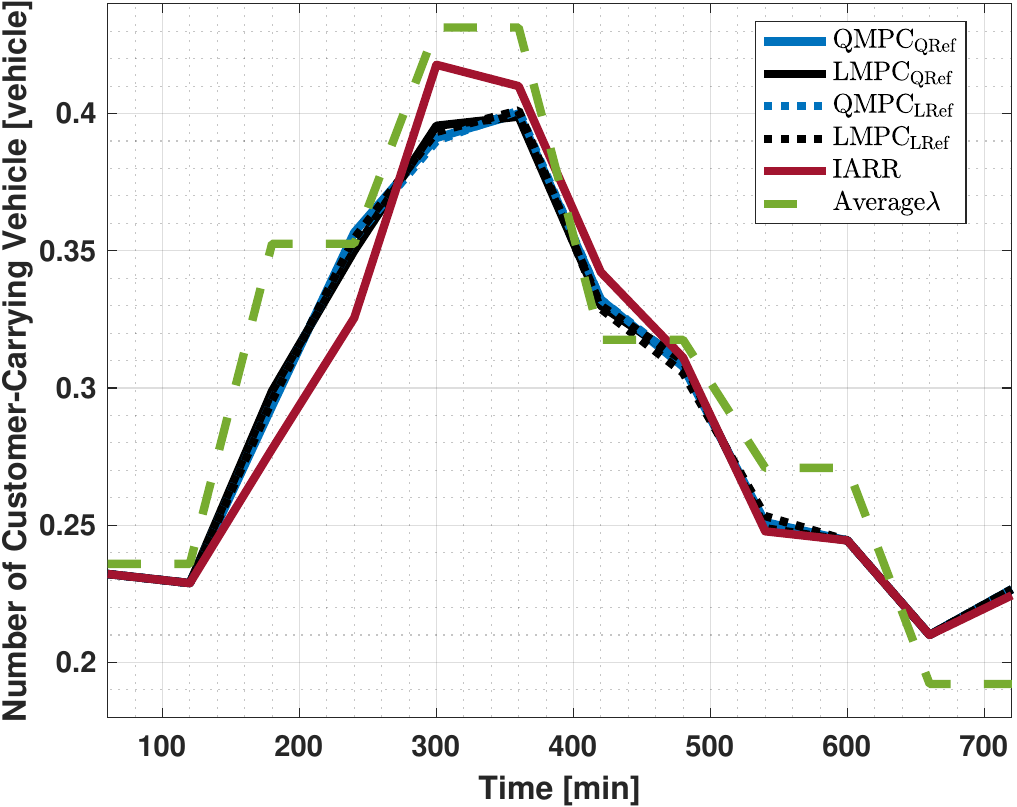}
 \caption{Case Study 1: The average number of customer-carrying vehicles.}
   \label{fig: case1 U}
  \end{subfigure}
  \hfill
\begin{subfigure}{0.48\textwidth}
    \centering
  \includegraphics[width=1\textwidth]{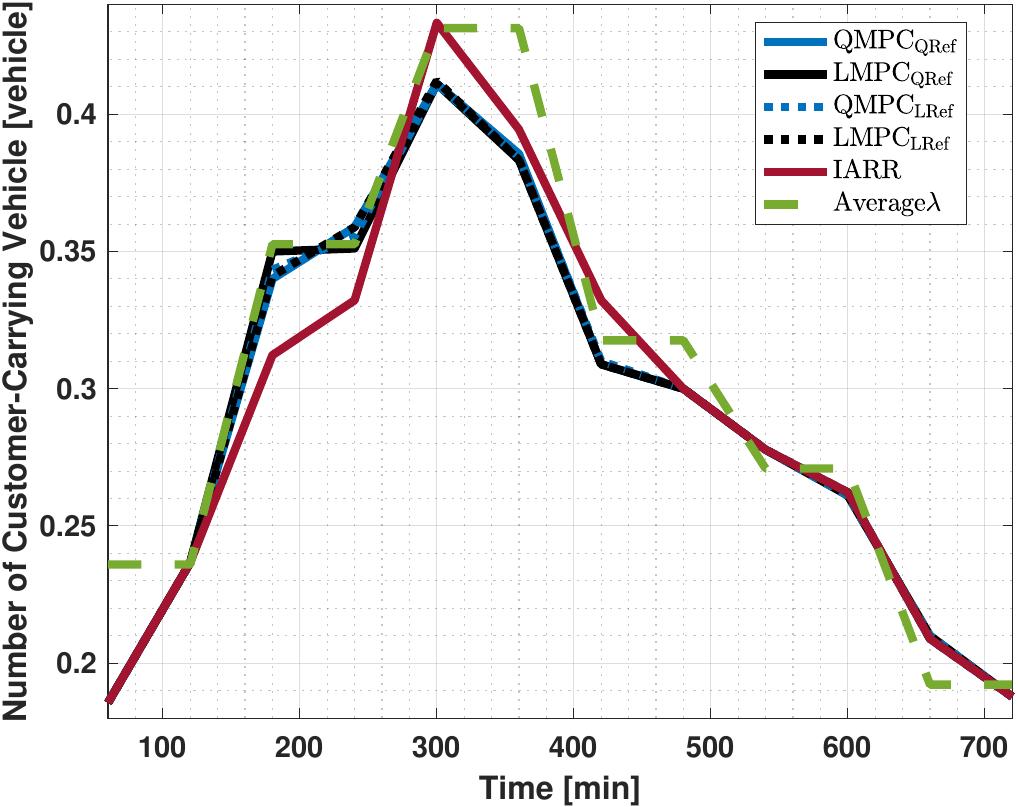}
 \caption{Case Study 2: The average number of customer-carrying vehicles.}
   \label{fig: case2 U}
  \end{subfigure}
  	\caption{The average queue length and number of customer-carrying vehicles for Case Studies 1 and 2.}
  	  \label{fig: Convergence}
  	\end{figure*}
Figures \ref{fig: case1 W} and \ref{fig: case2 W} display the average queue length for all origin-destination pairs in Case Studies 1 and 2, respectively. In the first 120 minutes, the queue length remains constant in Case Study 1 but increases for all algorithms in Case Study 2. Similar increases are observed, with the most notable one occurring in the IARR for both case studies, extending to the next 120 minutes. Following the demand pattern in Fig. \ref{fig:demand_histogram}, after 300 minutes, all algorithms exhibit a decreasing trend, and they display nearly identical behavior after 480 minutes.

Figures \ref{fig: case1 U} and \ref{fig: case2 U} show the average number of customer-carrying vehicles per link for Case Studies 1 and 2. While the overall trend in the average customer-carrying vehicles aligns with the pattern in Fig. \ref{fig:demand_histogram}, MPC-based algorithms dispatch a higher number of customer-carrying vehicles in the interval between 120 and 270 minutes. Subsequently, from 270 minutes to 480 minutes, the IARR algorithm takes the lead in terms of customer-carrying vehicles, and beyond that point, all algorithms exhibit nearly identical behavior.

\begin{figure*}[thbp]
  \begin{subfigure}{0.48\textwidth}
    \centering
  \includegraphics[width=1\textwidth]{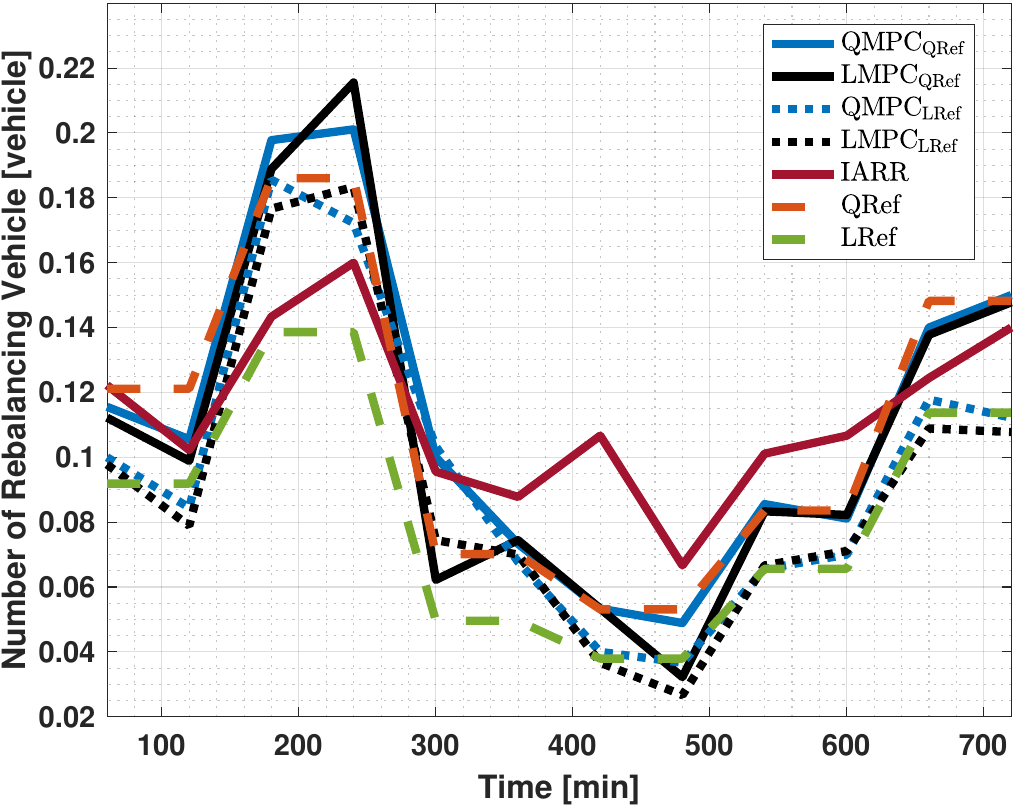}
 \caption{Case Study 1: The average number of rebalancing vehicles.}
   \label{fig: case1 R}
  \end{subfigure}
  \hfill
\begin{subfigure}{0.48\textwidth}
    \centering
  \includegraphics[width=1\textwidth]{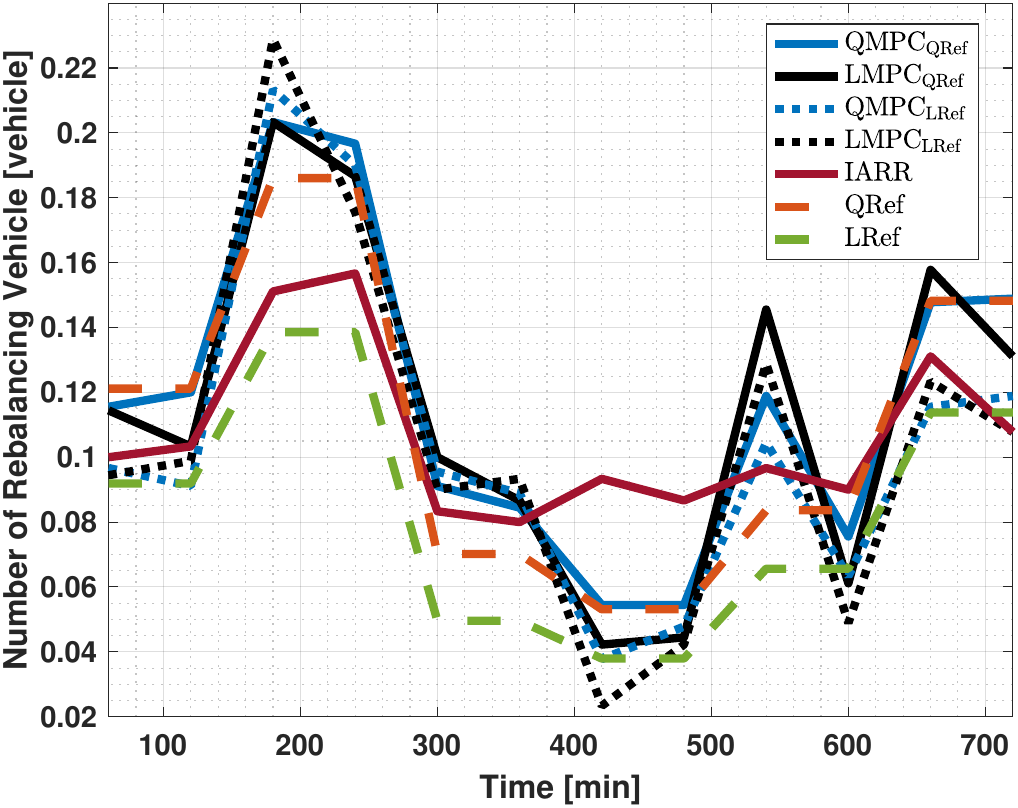}
 \caption{Case Study 2: The average number of rebalancing vehicles.}
   \label{fig: case2 R}
  \end{subfigure}  

    \begin{subfigure}{0.48\textwidth}
    \centering
  \includegraphics[width=1\textwidth]{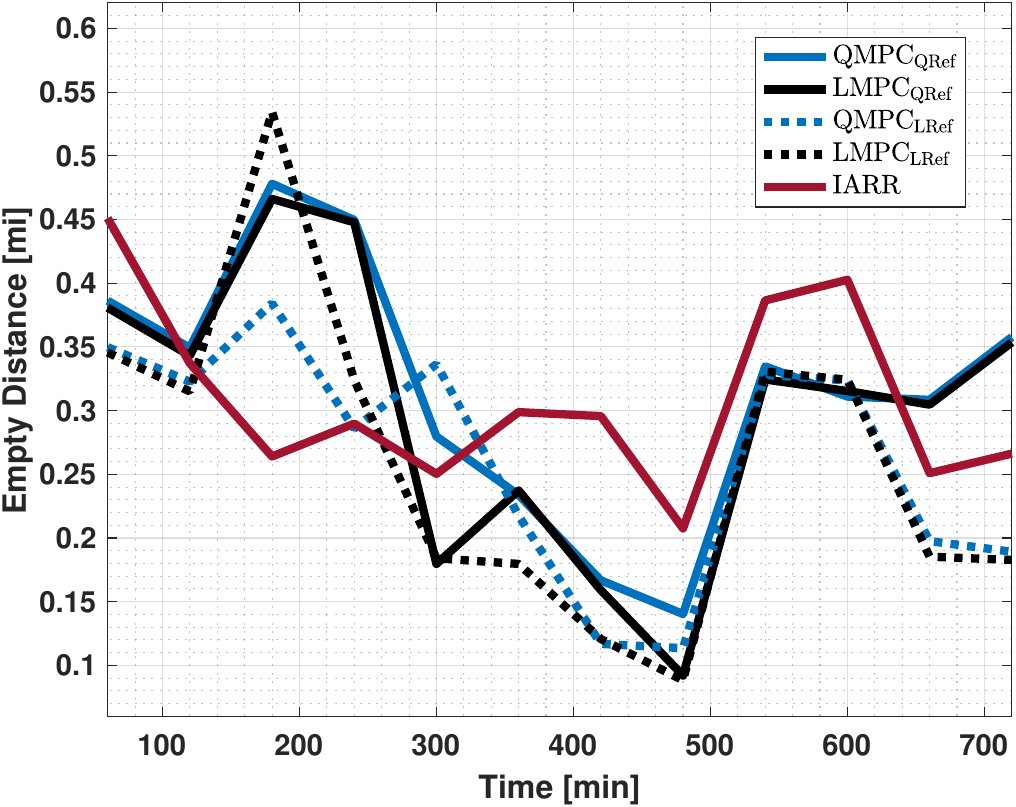}
  
   \caption{Case Study 1: The average empty distance.}
   \label{fig: case1 ED}
  \end{subfigure}
  \hfill
\begin{subfigure}{0.48\textwidth}
    \centering
  \includegraphics[width=1\textwidth]{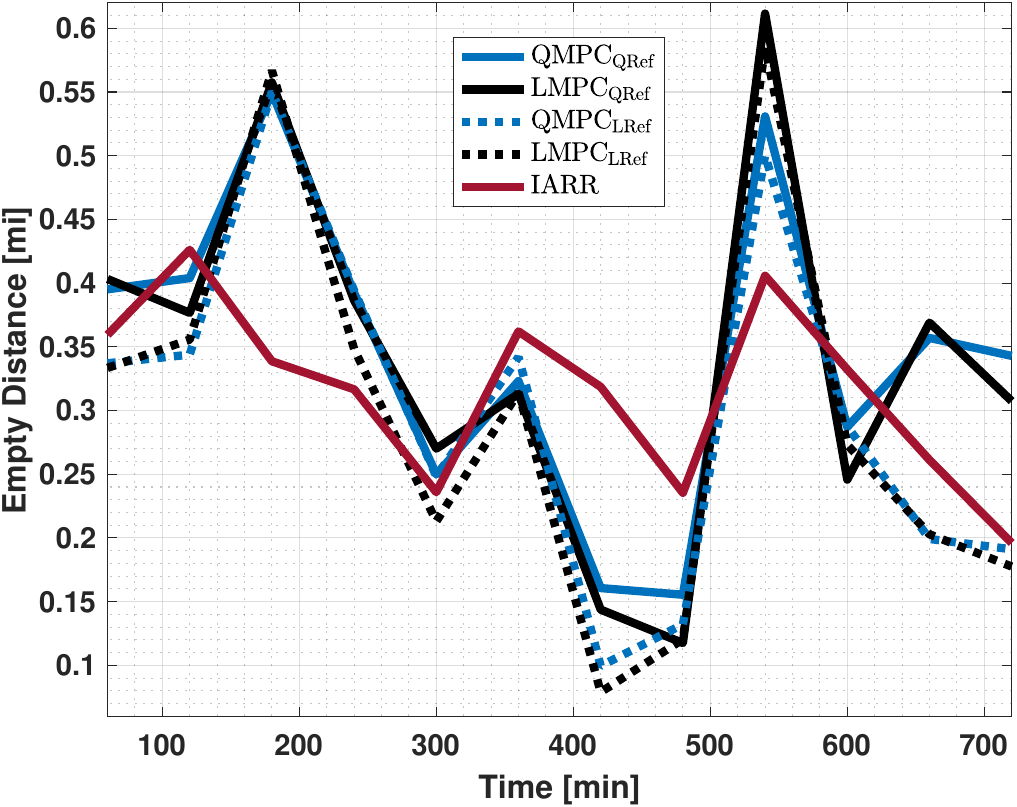}
   \caption{Case Study 2: The average empty distance.}
   \label{fig: case2 ED}
  \end{subfigure}  
  	\caption{The average number of rebalancing vehicles and empty distance for Case Studies 1 and 2.}
  	  \label{fig: Convergence}
  	\end{figure*}

Figures \ref{fig: case1 R} and \ref{fig: case2 R} illustrate the average number of rebalancing vehicles for Case Studies 1 and 2, respectively. Additionally, Figures \ref{fig: case1 ED} and \ref{fig: case2 ED} show the distance driven by the rebalancing vehicles in the same case studies.
As observed in Figures \ref{fig: case1 R} and \ref{fig: case2 R}, MPC-based algorithms demonstrate a higher level of rebalancing and more efficient handling of peak hour scenarios (interval between 240 and 360 minutes), irrespective of the tracked reference. As we move beyond the peak hour and experience a decrease in demand, the need for rebalancing decreases as well. However, the IARR algorithm performs rebalancing with less sensitivity to the demand changes. This predictability is one of the strengths of the MPC approach.

   \begin{figure*}[thbp]
  \begin{subfigure}{0.48\textwidth}
    \centering
  \includegraphics[width=1\textwidth]{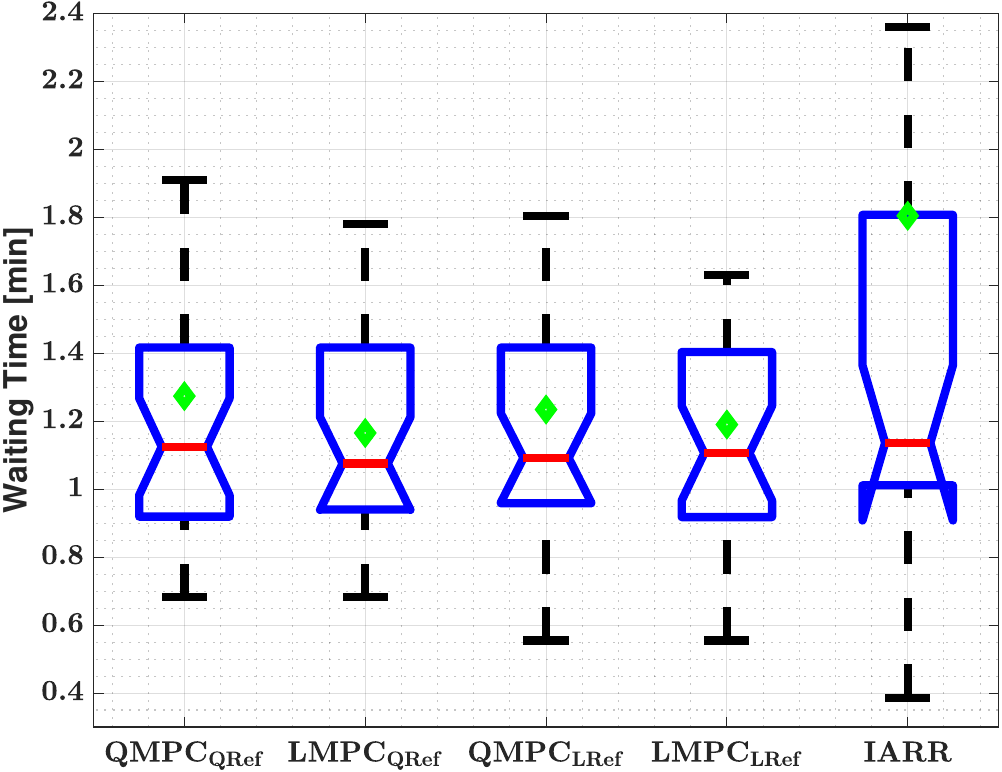}
 \caption{Case Study 1: The average customer waiting time at origin.}
   \label{fig: case1 theta}
  \end{subfigure}
  \hfill
\begin{subfigure}{0.48\textwidth}
    \centering
  \includegraphics[width=1\textwidth]{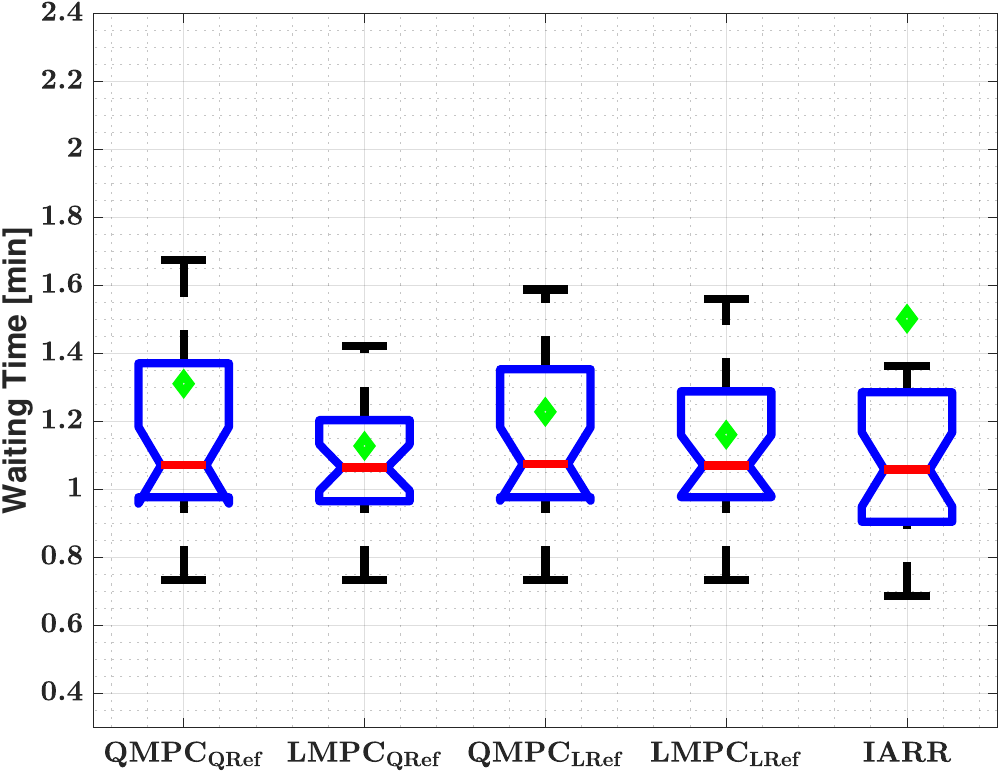}
 \caption{Case Study 2: The average customer waiting time at origin.}

   \label{fig: case2 theta}
  \end{subfigure}  
  \caption{A box plot showing the variations of the average waiting time for different scenarios. Red markers signify the median, while green markers indicate the average.}
  	  \label{fig: Convergence}
  	\end{figure*}

Figures \ref{fig: case1 theta} and \ref{fig: case2 theta} depict box plots showing waiting times for all origin-destination pairs. In both case studies, the median and average waiting times are similar, with slightly lower values for LMPC. Notably, the interquartile ranges in Case Study 2 are generally lower than in Case Study 2. Furthermore, in Case Study 1, the IARR algorithm exhibits a significantly higher interquartile range compared to MPC-based algorithms.

\begin{table*}[thbp]
	\caption{The average waiting customer and time as well as the distances with the empty vehicles during the 12-h simulation.}
	\label{tab:tableEQ}
	\begin{center}
				\begin{tabular}{l l l l l l}
			\hline
			&QMPC$_{\rm QRef}$ &QMPC$_{\rm LRef}$& LMPC$_{\rm QRef}$&LMPC$_{\rm LRef}$&IARR \\
\hline
\hline
{\bf Case study 1}  & & &&& \\
Average queue length & 0.6103 & 0.6059 & 0.5971& 0.5996 & 0.8468 \\
Average waiting time [min] & 1.2738 & 1.2343 &  1.1653 &1.1901&1.8045 \\
Total empty distance [mi] & 3,416 & 2,849 & 3,243& 2,803& 3,330\\
\hline
{\bf Case study 2}  & & &&& \\
Average queue length & 0.6116 & 0.6108 & 0.5989 & 0.6059& 0.8147\\
Average waiting time [min] & 1.3109 & 1.2282 & 1.1278 &1.1610&1.5021 \\
Total empty distance [mi] & 3,733 & 3,263 & 3,692&3,212& 3,409\\
\hline
\end{tabular}
\end{center}
\end{table*}
Table~\ref{tab:tableEQ} shows the average queue length and waiting time as well as the distances driven by empty vehicles during the 12-h simulation. 
In both case studies, all MPC-based algorithms exhibit similar average queue lengths, with slight differences favoring LMPC algorithms. However, the difference in average waiting times is more significant, and all MPC-based algorithms outperform IARR by approximately 28\%. Additionally, for both case studies, MPC algorithms with linear references have lower total empty distances. This is expected because the average linear reference is lower than the average quadratic reference being tracked. IARR's total empty distance is more comparable to MPCs with quadratic references for rebalancing. In the IARR algorithm, the primary focus is on minimizing the number of rebalancing vehicles. When considering the empty distance measure, QMPC${\rm LRef}$ and LMPC${\rm LRef}$ demonstrate superior performance. In Case Study 2, the total empty distance for all algorithms increased, although the average waiting time slightly decreased, except for QMPC$_{\rm QRef}$. Therefore, if maintaining a high quality of service is a critical goal, Case Study 2 proves superiority. This suggests that vehicle dispatching should occur in smaller time intervals. However, for a comprehensive evaluation, one should consider both the quality of service and the performance level, in which Case Study 2 is dominant.

The numerical results indicate that LMPC algorithms outperform QMPC algorithms, in terms of lower average waiting times and queue lengths. In Case Study 1, LMPC$_{\rm QRef}$ achieves a 2.1\% reduction in waiting time at the expense of a 15.7\% increase in total empty distances compared to LMPC$_{\rm LRef}$. In Case Study 2, LMPC$_{\rm QRef}$ demonstrates a 2.8\% lower waiting time with a 14.9\% increase in total empty distances compared to LMPC$_{\rm QRef}$. Considering all the evaluation criteria, it is evident that LMPC$_{\rm LRef}$ proves to be effective and outperforms other MPC-based and IARR algorithms, as shown in Table~\ref{tab:tableEQ}.
MPC requires state feedback from the plant, \cite{aalipour2018analytical}, every control step (which is set to 2 [min] and 3 [min] for Case studies 1 and 2, respectively). Considering this inherent structure and the rolling-horizon feature of MPC make it more robust by design and allow it to tackle modeling errors and hinder the accumulation of errors in states over time.

\section{Conclusion}
\label{sec: conclusion}
In this paper, we proposed a linear time-delay model for the AMoD system. After introducing an approximation of the model, we discussed the basic properties of the model. We presented an optimization problem to derive the optimal rebalancing reference being tracked by MPC in the AMoD system.
Afterward, we applied the MPC framework for regulation around the optimal equilibrium point to derive the optimal rebalancing policy.
In summary, if the minimum required fleet size is established and the AMoD system is properly managed, it is unlikely to run into an imbalanced system. In addition, the simulation results showed the performance and properties of the MPC algorithm. 
Numerical results on the University of Minnesota's campus network show that if both the MPC cost function and optimal rebalancing cost function are linear, it outperforms the other MPC-based and IARR algorithms according to the quality of service to customers and empty distance driven by the rebalancing vehicles.

Future work involves investigating the scalability of the proposed model to a larger network. Besides that, developing model-free reinforcement learning algorithms would be another direction for future work. In addition,  developing a model for electric vehicles and connecting AMoD to public transit to improve first/last mile access is another exciting work to consider.
\section*{Acknowledgement}
This research is conducted at the University of Minnesota Transit Lab, currently supported by the following, but not limited to, projects:
\begin{itemize}
     \item[--] National Science Foundation, award CMMI-1831140
     \item[--] Freight Mobility Research Institute (FMRI), Tier 1 Transportation Center, U.S. Department of Transportation
     \item[--] Minnesota Department of Transportation
\end{itemize}

\appendix

\section{Proof of Theorem \ref{theorem:MPC_convergence}}
\label{Theorem: Proof of Main Theorem}
\begin{Theorem}[Lyapunov Drift Theorem (\cite{neely2022stochastic})]
Consider the Lyapunov function 
$$L({\bf W}(t)) = {\frac {1}{2}}\sum_{(r,s) \in N^2}^{}{W}_{rs}(t)^{2}.$$ Suppose there exists a value function $L({\bf W}(t))$, and there are constants $B < \infty$ and $\varepsilon > 0 $ such that for all $t$ and all possible vectors ${\bf W}(t)$ the conditional Lyapunov drift satisfies: 
$\mathbb {E} [ L({\bf W}(t+1) - L({\bf W}(t)))|{\bf W}(t)]\leq B - \varepsilon \|{\bf W}(t)\|_1$, then for all slots $t > 0$, the time average queue size in the network satisfies:
\begin{align}
{\frac {1}{t}}\sum_{\tau =0}^{t-1}\sum _{(r,s) \in N^2}^{}\mathbb {E} [{W}_{rs}(\tau )]\leq {\frac {B}{\varepsilon }}+{\frac {\mathbb {E} [L(0)]}{\varepsilon t}}.
\label{Th: telesc}
\end{align}
\end{Theorem}

Let $D = \{{\bf d}: \forall {\bf U}(t) \in {\mathcal U} \text{ s.t. } {\bf U}(t) \geq \lambda \}$ and $D^{0}$ be the interior of $D$. Now, if ${\bf d} \in D^{0}$ and the dispatch policy uses average flow ${\bf V}$, there exists an 
 $B < \infty$ and $\varepsilon > 0 $ such that for all $t$ and all possible vectors ${\bf W}(t)$, the conditional Lyapunov drift satisfies: 
\begin{align}
\mathbb {E} [ L({\bf W}(t+1) - L({\bf W}(t)))|{\bf W}(t)]\leq B - \varepsilon \|{\bf W}(t)\|_1.
\label{eq: Lya drift cond}
\end{align}
This can be shown by taking the expectation over \eqref{eq: Lya drift cond}, and summing over all the $t$. Finally, using the telescoping properties, Eq. \eqref{Th: telesc} can be derived.

Since $V_{rs}(t) = \min \{W_{rs}(t),U_{rs}(t)\}$, if $W_{rs}(t) < U_{rs}(t)$, then $V_{rs}(t) = W_{rs}(t)$ and
\begin{align*}
\mathbb {E} [ \Delta L({\bf W}(t))|{\bf W}(t)] &= \mathbb {E} [ L({\bf W}(t+1) - L({\bf W}(t))|{\bf W}(t)] 
\\ &=  {\frac {1}{2}} \sum_{(r,s) \in N^2}^{}
{\lambda}^2_{rs} + {\lambda}_{rs} - W^2_{rs}(t) \leq {\frac {1}{2}} \sum_{(r,s) \in N^2}^{}
{\lambda}^2_{rs} + {\lambda}_{rs} - {\frac {1}{2}} \|{\bf W}(t)\|_1.
\end{align*}
If $W_{rs}(t) \geq U_{rs}(t)$, then $V_{rs}(t) = U_{rs}(t)$ and 
\begin{align*}
 \mathbb {E} [ \Delta L({\bf W}(t))|{\bf W}(t)] =& \mathbb {E} [ L({\bf W}(t+1)) - L({\bf W}(t))|{\bf W}(t)] = {\frac {1}{2}}  \mathbb {E} [\sum_{(r,s) \in N^2}^{}{W}_{rs}(t+1)^{2} - {W}_{rs}(t)^{2} |{\bf W}(t)] 
\\ = &  \mathbb {E} [\sum_{(r,s) \in N^2}^{} {W}_{rs}(t)(d_{rs}(t) - U_{rs}(t)) + {\frac {(d_{rs}(t) - U_{rs}(t))^2}{2}} |{\bf W}(t)].
\end{align*}
Since ${\bf d} \in D^0$, $\lambda - {\bf U}(t) < -\epsilon$, 
$\lambda_{rs} - U_{rs} (t) < -\epsilon$, $\forall (r,s) \in N ^2$. Hence, $-2\lambda_{rs}U_{rs} (t)\leq -\lambda_{rs}^2 + \lambda_{rs}(1-2\epsilon)$. Moreover:
\begin{align*}
 \frac{1}{2} \mathbb {E} [\sum_{(r,s) \in N^2}^{} (d_{rs}(t) - U_{rs}(t))^2 |{\bf W}(t)] &= 
\frac{1}{2}  \sum_{(r,s) \in N^2}^{} ({\lambda}^2_{rs} + {\lambda}_{rs} +{U}^2_{rs}(t)  - 2{\lambda}_{rs}{U}_{rs}(t) )\\
& \leq \frac{M^2}{2}  + \sum_{(r,s) \in N^2}^{} 2( {\lambda}_{rs} - \epsilon) = C_1.
\end{align*}
Consequently,
\begin{align*}
& \mathbb {E} [ \Delta L({\bf W}(t))|{\bf W}(t)] \leq C_1 - \|{\bf W}(t)\|_1.
\end{align*}

When the MPC policy is used with ${\bf d} \in D^0$, there exists an $N_{\rm hor}$ such that for all (lookahead) time horizons
$T > N_{\rm hor}$,
\begin{align*}
\mathbb {E} [\frac{1}{T}\sum_{\tau = 1}^{T}\Delta L({\bf W}(t+\tau))|{\bf W}(t)] \leq B - \epsilon \|{\bf W}(t)\|_1.  
\end{align*}
Expanding it will result in:
\begin{align*}
 \mathbb {E}& [\frac{1}{T}\sum_{\tau = 1}^{T}\sum_{(r,s) \in N^2}^{} \frac{W^2_{rs}(t+\tau + 1) - W^2_{rs}(t+\tau)}{2}|{\bf W}(t)] \\
 \leq&
\mathbb {E} [\frac{1}{T}\sum_{\tau = 1}^{T}\sum_{(r,s) \in N^2}^{}{W}_{rs}(t+\tau) (d_{rs}(t+\tau) -V_{rs}(t+\tau))|{\bf W}(t)] \\
&+\mathbb {E} [\frac{1}{T}\sum_{\tau = 1}^{T}\sum_{(r,s) \in N^2}^{} \frac{(d_{rs}(t+\tau) -V_{rs}(t+\tau))^2}{2}|{\bf W}(t)].
\end{align*}
We showed earlier that the second term on the right-hand side of the above inequality is upper bounded by $C_1$. Now, we are left with the first term of the right-hand side of the above inequality.
Also, given ${\bf W}(t)$, ${W}_{rs}(t+\tau)$ inside the expectation can be written as $$
  {W}_{rs}(t+\tau)= W_{rs}(t) + \tau \lambda_{rs} - \sum_{\tau^{'} = 0}^{\tau - 1} {V_{rs}(t+\tau^{'})},$$
and consequently,
\begin{align*}
\mathbb {E} &[\frac{1}{T}\sum_{\tau = 1}^{T}\sum_{(r,s) \in N^2}^{}{W}_{rs}(t+\tau) (d_{rs}(t+\tau) -V_{rs}(t+\tau))|{\bf W}(t)] \\
 &= \frac{1}{T}\sum_{\tau = 1}^{T}\sum_{(r,s) \in N^2}^{}(W_{rs}(t) + \tau \lambda_{rs}) (\lambda_{rs} -V_{rs}(t+\tau)) - \frac{1}{T}\sum_{\tau = 1}^{T}\sum_{(r,s) \in N^2}^{}(\sum_{\tau^{'} = 0}^{\tau - 1} {V_{rs}(t+\tau^{'})}) (\lambda_{rs} -V_{rs}(t+\tau)).
\end{align*}
Since $\sum_{r,s\in N^2}^{}{V_{rs}(t+\tau)}$ and $\sum_{r,s\in N^2}^{}{V_{rs}(t+\tau^{'})}$ are between zero and fleet size $M$,
$$
- \frac{1}{T}\sum_{\tau = 1}^{T}\sum_{(r,s) \in N^2}^{}(\sum_{\tau^{'} = 0}^{\tau - 1} {V_{rs}(t+\tau^{'})}) (\lambda_{rs} -V_{rs}(t+\tau))
$$
can be upper-bounded by a constant $C_2$.
Hence,
\begin{align*}
\frac{1}{T}\sum_{\tau = 1}^{T}\sum_{(r,s) \in N^2}^{} (W_{rs}(t) + \tau \lambda_{rs})(\lambda_{rs} -V_{rs}(t+\tau)) =& \frac{1}{T}\sum_{\tau = 1}^{T}\sum_{(r,s) \in N^2}^{}W_{rs}(t) (\lambda_{rs} -V_{rs}(t+\tau)) 
\\ &+ \frac{1}{T}\sum_{\tau = 1}^{T}\sum_{(r,s) \in N^2}^{}\tau \lambda_{rs} (\lambda_{rs} -V_{rs}(t+\tau)).
\end{align*}
The second term of the right-hand side can be bounded by $\frac{T+1}{2}\sum_{(r,s) \in N^2}^{}\lambda^2_{rs} = C_3$; thus,
we are left with the first term of the right-hand side, which is as follows:
\begin{align*}
& \frac{1}{T}\sum_{\tau = 1}^{T}\sum_{(r,s) \in N^2}^{}W_{rs}(t) (\lambda_{rs} -V_{rs}(t+\tau)).   
\end{align*}

Depending on the objective function in Problem \ref{Main_optimization}, two cases can be considered: quadratic and linear objective functions.
\begin{itemize}
    \item{Quadratic objective function:} In this case, by solving Problem \ref{Main_optimization}, the following equation can be derived:
\begin{align*}
\sum_{\tau = 1}^{T}\sum_{(r,s) \in N^2}^{} W^2_{rs}(t+\tau+1) |_{U^{*}}
\leq \sum_{\tau = 1}^{T}\sum_{(r,s) \in N^2}^{} W^2_{rs}(t+\tau+1) |_{U}.
\end{align*}
Expanding it will result in:
\begin{align*}
\sum_{\tau = 1}^{T}\sum_{(r,s) \in N^2}^{}
(W_{rs}(t+\tau) +\lambda_{rs} - V^{*}_{rs}(t + \tau))^2
\leq \sum_{\tau = 1}^{T}\sum_{(r,s) \in N^2}^{}
(W_{rs}(t+\tau) +\lambda_{rs} - V_{rs}(t + \tau))^2.
\end{align*}
Considering $W_{rs}(t+\tau) = W_{rs}(t) + \tau \lambda_{rs} - \sum_{\tau^{'} = 0}^{\tau}V_{rs}(t+\tau^{'})$.
Therefore,
\begin{align*}
\sum_{\tau = 1}^{T}\sum_{(r,s) \in N^2}^{}
W_{rs}(t)(\lambda_{rs} - V^{*}_{rs}(t + \tau)) + C_4
\leq \sum_{\tau = 1}^{T}\sum_{(r,s) \in N^2}^{}
W_{rs}(t)(\lambda_{rs} - V_{rs}(t + \tau)).
\end{align*}
By adding and subtracting $U^{*}_{rs}(t + \tau)$ on the left hand side, and $U_{rs}(t + \tau)$ on the right hand side, we have:
\begin{align*}
&\sum_{\tau = 1}^{T}\sum_{(r,s) \in N^2}^{}
W_{rs}(t)(\lambda_{rs} -U^{*}_{rs}(t + \tau)+ U^{*}_{rs}(t + \tau) - V^{*}_{rs}(t + \tau)) \\
&\leq  \sum_{\tau = 1}^{T}\sum_{(r,s) \in N^2}^{}
W_{rs}(t)(\lambda_{rs} -U_{rs}(t + \tau) U_{rs}(t + \tau)- V_{rs}(t + \tau))+C_4.
\end{align*}
The added terms are bounded by a constant because if $W_{rs}(t+\tau) \geq U_{rs}(t+\tau)$, then
$U_{rs}(t+\tau) - V_{rs}(t+\tau) = 0$. On the other hand, if $W_{rs}(t+\tau) \leq U_{rs}(t+\tau)$, $W_{rs}(t)$ can be upper bounded by $U_{rs}(t+\tau)$ and $U_{rs}(t+\tau) - V_{rs}(t+\tau)$ can be upper bounded by $U_{rs}(t+\tau)$. $U_{rs}(t+\tau)$ is upper bounded by the fleet size $M$, which yields

\begin{align*}
\sum_{\tau = 1}^{T}\sum_{(r,s) \in N^2}^{}
W_{rs}(t)(\lambda_{rs} -U_{rs}(t + \tau)+ U_{rs}(t + \tau) - V_{rs}(t + \tau)) 
\leq  T (\sum_{(r,s) \in N^2}^{}
U_{rs}(t))^2\leq T M^2.
\end{align*}

Therefore, we can conclude the following
\begin{align*}
\sum_{\tau = 1}^{T}\sum_{(r,s) \in N^2}^{}
W_{rs}(t)(\lambda_{rs} -U^{*}_{rs}(t + \tau))\leq& \sum_{\tau = 1}^{T}\sum_{(r,s) \in N^2}^{}
W_{rs}(t)(\lambda_{rs} -U_{rs}(t + \tau)) + C_5\\
\leq& -\epsilon\sum_{\tau = 1}^{T}\sum_{(r,s) \in N^2}^{}
W_{rs}(t) + C_5\\
=&C_5 -T\epsilon \|{\bf W}(t)\|_1,
\end{align*}
where $C_5 = C_4 + TM^2$.

\item{Linear objective function:}
\begin{align*}
\sum_{\tau = 1}^{T}\sum_{(r,s) \in N^2}^{} W_{rs}(t+\tau+1) |_{U^{*}}
\leq \sum_{\tau = 1}^{T}
\sum_{(r,s) \in N^2}^{} W_{rs}(t+\tau+1) |_{U}.
\end{align*}

Since ${\bf W}(t) \geq 0$, using 
$$
\| {\bf W} \|_2^2 \leq \| {\bf W} \|_1^2\leq {s_w}\| {\bf W} \|_2^2,
$$
where $s_w$ is the size of the vector ${\bf W}$. Hence, we can have
\begin{align*}
\sum_{\tau = 1}^{T}\sum_{(r,s) \in N^2}^{} W^2_{rs}(t+\tau+1) |_{U^{*}}&\leq {s_w} \sum_{\tau = 1}^{T}
\sum_{(r,s) \in N^2}^{} W^2_{rs}(t+\tau+1) |_{U}
\\
& \leq  {s_w}\left(C_5 -  T\epsilon\sum_{\tau = 1}^{T}\sum_{(r,s) \in N^2}^{}
W_{rs}(t) \right)\\
&={s_w}C_5  - {s_w} \epsilon T\|{\bf W}(t)\|_1.
\end{align*}

\end{itemize}

\bibliographystyle{ieeetr}
\bibliography{main}

\end{document}